\documentstyle[12pt]{article}

\font\blackboard=msbm10 at 12pt
\font\blackboards=msbm7
\font\blackboardss=msbm5
\newfam\black
\textfont\black=\blackboard
\scriptfont\black=\blackboards
\scriptscriptfont\black=\blackboardss
\def\bb#1{{\fam\black\relax#1}}

\newcommand{\junk}[1]{}

\newcommand{\ba}{\begin{array}}
\newcommand{\ea}{\end{array}}
\newcommand{\be}{\begin{equation}}
\newcommand{\ee}{\end{equation}}
\newcommand{\bea}{\begin{eqnarray}}
\newcommand{\eea}{\end{eqnarray}}
\newcommand{\beas}{\begin{eqnarray*}}
\newcommand{\eeas}{\end{eqnarray*}}

\def\identity{{\rlap{1} \hskip 1.6pt \hbox{1}}}

\def\laplace{{\kern1pt\vbox{\hrule height 1.2pt\hbox{\vrule width
1.2pt\hskip
  3pt\vbox{\vskip 6pt}\hskip 3pt\vrule width 0.6pt}\hrule height
  0.6pt}
  \kern1pt}}
\def\scriptlap{{\kern1pt\vbox{\hrule height 0.8pt\hbox{\vrule width
  0.8pt
  \hskip2pt\vbox{\vskip 4pt}\hskip 2pt\vrule width 0.4pt}\hrule height
  0.4pt}
  \kern1pt}}
\def\slash#1{{\rlap{$#1$} \thinspace /}}
\def\roughly#1{\raise.3ex\hbox{$#1$\kern-
.75em\lower1ex\hbox{$\sim$}}}

\def\str{{\rm STr} \,}
\def\sym{{\rm Sym} \,}

\def\tr{{\rm Tr} \,}

\def\ab{{\bar{\alpha}}}
\def\bb{{\bar{\beta}}}

\def\ha{{\hat{a}}}

\newcommand{\NP}{{\em Nucl.\ Phys.\ }}

\newcommand{\PL}{{\em Phys.\ Lett.\ }}
\newcommand{\PR}{{\em Phys.\ Rev.\ }}

\newcommand{\MPL}{{\em Mod.\ Phys.\ Lett.\ }}
\newcommand{\PRL}{{\em Phys.\ Rev.\ Lett.\ }}

\newcommand{\gone}[1]{}
\begin{document}
\pagestyle{plain}
\setcounter{page}{1}

\baselineskip16pt

\begin{titlepage}

\begin{flushright}
PUPT-1894\\
MIT-CTP-2905\\
hep-th/9910052
\end{flushright}
\vspace{8 mm}

\begin{center}

{\Large \bf Multiple D$p$-branes in Weak Background Fields\\}

\end{center}

\vspace{7 mm}

\begin{center}

Washington Taylor IV$^a$ and Mark Van Raamsdonk$^b$

\vspace{3mm}
${}^a${\small \sl Center for Theoretical Physics} \\
{\small \sl MIT, Bldg.  6-306} \\
{\small \sl Cambridge, MA 02139, U.S.A.} \\
{\small \tt wati@mit.edu}\\

\vspace{3mm}
${}^b${\small \sl Department of Physics} \\
{\small \sl Joseph Henry Laboratories} \\
{\small \sl Princeton University} \\
{\small \sl Princeton, New Jersey 08544, U.S.A.} \\
{\small \tt mav@princeton.edu}
\end{center}

\vspace{8 mm}

\begin{abstract}
We find the terms in the nonabelian world-volume action of a system of
many D$p$-branes which describe the leading coupling to all type II
supergravity background fields.  These results are found by
T-dualizing earlier results for D0-branes, which in turn were
determined from calculations of the M(atrix) theory description of the
supercurrent of 11D supergravity.  Our results are compatible with
earlier results on the supersymmetric Born-Infeld action for a single
D-brane in a general background and with Tseytlin's symmetrized trace
proposal for extending the abelian Born-Infeld action to a nonabelian
theory. In the case $p = 3$, the operators we find on the D-brane
world-volume are closely related to those which couple to supergravity
fields in the AdS${}_5\times S^5$ IIB supergravity background.  This
gives an explicit construction, including normalization, of some of
the operators used in the celebrated AdS/CFT correspondence for
3-branes.  We also discuss the S-duality of the action in the case $p
= 3$, finding that the S-duality of the action determines how certain
operators in the ${\cal N} = 4$ 4D SYM theory transform under
S-duality.  These S-duality results give some new insight into the
puzzle of the transverse 5-brane in M(atrix) theory.
\end{abstract}

\vspace{1cm}
\begin{flushleft}
October 1999
\end{flushleft}
\end{titlepage}
\newpage

\section{Introduction}

Since Polchinski's realization in 1995 \cite{Polchinski} that the
Dirichlet $p$-branes of string theory carry Ramond-Ramond (R-R) charge
and should be identified with R-R charged black brane solutions of
supergravity, D-branes have become one of the most important tools in
analyzing string theory and M-theory.  D-branes have been used to
construct and study black holes and supersymmetric field theories with
many interesting properties.  It is even believed that D-branes can be
used to describe all of string theory and M-theory in certain regimes
and in certain backgrounds, through the M(atrix) theory \cite{BFSS}
and AdS/CFT proposals \cite{Maldacena-conjecture}.  (For reviews and
further references see \cite{Polchinski-TASI,WT-Trieste} for general
background on D-branes, \cite{Peet-review} for black holes from
D-branes, \cite{Giveon-Kutasov} for field theories from D-branes,
\cite{banks-review,Susskind-review,Bilal-review,WT-Trieste} for
M(atrix) theory and \cite{agmoo} for the AdS/CFT correspondence.)

To understand the behavior of D-branes in different contexts, it is of
fundamental importance to have a description of their dynamics in a
general string theory background.  To date, our understanding of this
description is somewhat incomplete.  For a single D-brane it is known
that the classical dynamics in a general string background should have
a bosonic part described by the Born-Infeld action \cite{Leigh}
combined with Wess-Zumino terms describing the coupling to R-R
background fields \cite{Douglas} and to the curvature of the background
metric \cite{ghm,Minasian-Moore,Cheung-Yin}.  The fermionic part of the
theory can be included into a $\kappa$-symmetric action both in flat
space \cite{aps1,aps2} and in a general type II supergravity background
\cite{cgnw,cgnsw,Bergshoeff-Townsend}.  For a single Dirichlet $p$-brane
this story is essentially complete: the degrees of freedom in the
world-volume theory are the $9-p$ transverse scalar fields $X^i$, a
world-volume $U(1)$ gauge field $A_\alpha$ and fermionic fields which
complete a supersymmetry multiplet with $16$ supercharges.  In the
low-energy limit the supersymmetric Born-Infeld action for a
D$p$-brane reduces to the maximally supersymmetric $U(1)$ Yang-Mills
action in $p + 1$ dimensions. 

For a system of multiple D$p$-branes, the world-volume action is much
less well understood than for a single brane.  In the low-energy
limit, the action for a system of $N$ parallel D$p$-branes in a flat
supergravity background reduces to the maximally supersymmetric $U(N)$
Yang-Mills theory in $p + 1$ dimensions \cite{Witten-bound}.  The
extension of this action to a full supersymmetric or
$\kappa$-symmetric nonabelian Born-Infeld action is not known.  It was
suggested by Tseytlin \cite{Tseytlin} that at least for the bosonic
terms, such an extension can be found by using a symmetrized trace to
resolve ordering ambiguities inherent in the higher order terms of the
Born-Infeld action.  Although this proposal has not yet been derived
from any more fundamental principles, the symmetrized trace gives
results compatible with other approaches such as direct string
computation \cite{Gross-Witten,Tseytlin-vector} and M(atrix) theory
calculations \cite{Chepelev-Tseytlin,Dan-Wati-2}, at least at low
order in the field strength $F$ (see, however,
\cite{Hashimoto-Taylor,Bain} for a related puzzle).  A recent
review of the state of knowledge regarding both the
abelian and nonabelian Born-Infeld action is given in
\cite{Tseytlin-review}.

In this paper we find a new set of terms in the action describing a
system of multiple D$p$-branes in a general type II supergravity
background.  The new terms which we identify are terms in the action
which couple the world-volume fields linearly to the background
supergravity fields and all their higher derivatives.  For a given
derivative of a background field, we determine the lowest dimension
operator in the world-volume theory which couples to the background
field derivative at linear order.  In a previous paper
\cite{Mark-Wati-4} we described this set of couplings for a system of
multiple D0-branes.  These results were achieved by transforming
earlier results for matrix theory in a weak background
\cite{Mark-Wati-3} using the limiting procedure suggested by Seiberg
and Sen \cite{Seiberg-DLCQ,Sen-DLCQ} for relating the matrix
description of M-theory to  the $N$ D0-brane action in IIA
string theory.  In this paper we T-dualize the results of
\cite{Mark-Wati-4} to find the leading linear part of the nonabelian
Born-Infeld action for a D$p$-brane of arbitrary dimension. In addition, 
we show how T-duality may also be used starting with the known abelian 
D9-brane action to derive non-abelian terms in all lower dimensional
D$p$-brane 
actions, providing an alternate derivation of some of our results and 
giving a large set of additional terms.   

The paper is organized as follows. In Section 2 we review our results
for D0-branes and perform the T-duality transformation to find terms
in the action for a D-brane of arbitrary dimension. In Section 3 we
compare our results in the abelian case to the known
$\kappa$-symmetric action for a single D$p$-brane. We find that the
two methods are consistent, though there is a subtlety in the
comparison of fermion terms. We further show how the known abelian
D9-brane action may be used to derive a large set of additional terms
in the non-abelian actions for lower dimensional D$p$-branes using
T-duality. In Section 4 we discuss the S-duality of our action in the
case $p = 3$ and comment on a connection with the transverse 5-brane
of matrix theory. Section 5 contains some comments on the connection
of our results with the AdS/CFT correspondence, including a discussion
of how the actions we derive may be used in the study of D-brane black
holes.  We conclude in section 6 with a brief discussion of the
expected corrections to the actions we have derived as well as
possible further developments.

While we were completing this paper we received a preprint by Myers
\cite{Myers-dielectric} which contains some closely related results.  In
particular, the treatment in \cite{Myers-dielectric} is closely related
to our discussion in section 3.2.

\section{Linear brane-background couplings}

In this section we determine the lowest dimension operators on a
D$p$-brane of arbitrary dimension which couple linearly to the
derivatives of the type II background supergravity fields.  We assume
that the background fields represent small variations around a flat
space background.

\subsection{Review of D0-brane results}
\label{sec:review}

We begin by reviewing the method and results of our earlier paper 
\cite{Mark-Wati-4} describing D0-branes in weak background fields.

The main idea was to learn about D0-branes using matrix theory results 
by exploiting the relationship between the action for D0-branes in type 
IIA string theory and the matrix theory action which describes the DLCQ 
of M-theory. Given the action for D0-branes in a specified background, 
Seiberg  and Sen have given a very explicit prescription
\cite{Seiberg-DLCQ,Sen-DLCQ} for how to derive the 
matrix model of M-theory in the related background. For the case of flat 
space, the matrix theory action arises as the lowest dimension part of 
the string theory action describing the dynamics of D0-branes in flat 
space. Similarly, the action for matrix theory in a general 
11-dimensional 
supergravity background should arise from leading terms in 
the action for D0-branes in a general type IIA supergravity background. 
In previous work \cite{Mark-Wati-3}, we derived explicitly the action 
for matrix theory in the presence of arbitrary weak 11-dimensional 
supergravity fields, so our strategy in \cite{Mark-Wati-4} was to use 
these results and reverse the Seiberg-Sen prescription to deduce the 
leading 
terms in the D0-brane action in the presence of arbitrary type IIA 
supergravity fields.

In the general background matrix theory action, the 11-dimensional
supergravity background fields and their derivatives couple to matrix
theory expressions for the multipole moments of the various conserved
11-dimensional supergravity currents. Naturally, the background metric
couples to the matrix theory expressions for moments of the
stress-energy tensor, denoted by $T^{IJ(k_1\cdots k_n)}$ while the
three-form $A_{IJK}$ appears in the action coupled electrically to the
moments of the membrane current, $J^{IJK(k_1\cdots k_n)}$, and
magnetically to moments of the fivebrane current $M^{IJKLMN(k_1 \cdots
k_n)}$. The complete expressions for $T$, $J$, and $M$ were derived in
\cite{Dan-Wati-2,Mark-Wati-3} by comparing a matrix theory calculation
of the long-distance interaction potential between two completely
general systems with the analogous potential in supergravity. These
expressions are reproduced for convenience in the appendix.

Turning to the D0-brane action, the most general form for the terms 
linear in bosonic background fields can be written as
\begin{eqnarray}
S_{{\rm linear}}  &= &  S_{{\rm flat}}  +
 \int dt  \sum_{n=0}^{\infty} {1 
\over n!} \left[
\frac{1}{2}
(\partial_{k_1}\cdots 
\partial_{k_n} h^{IIA}_{\mu \nu}) \; I_h^{\mu \nu (k_1 \cdots k_n)}
+ (\partial_{k_1}\cdots \partial_{k_n} \phi) 
\; I_{\phi}^{(k_1 \cdots k_n)}\right.\label{eq:IIA-general}\\ 
& & \hspace{1in}+ (\partial_{k_1}\cdots 
\partial_{k_n} C_{\mu }) \; I_0^{\mu (k_1 \cdots k_n)}
+
 (\partial_{k_1}\cdots 
\partial_{k_n} \tilde{C}_{\mu \nu \lambda \rho \sigma \tau \zeta }) 
\; I_6^{\mu \nu \lambda
\rho \sigma \tau \zeta (k_1 \cdots k_n)}
\nonumber\\
& & \hspace{1in}+
 (\partial_{k_1}\cdots \partial_{k_n} 
B_{\mu \nu}) \; I_s^{\mu \nu (k_1 \cdots k_n)} 
+ 
(\partial_{k_1}\cdots \partial_{k_n} 
\tilde{B}_{\mu \nu \lambda \rho \sigma \tau}) \; I_5^{\mu \nu \lambda
\rho \sigma \tau
 (k_1 \cdots k_n)} \nonumber\\
& &\hspace{1in} \left.+ 
(\partial_{k_1}\cdots 
\partial_{k_n} C^{(3)}_{\mu \nu \lambda }) \; I_2^{\mu \nu \lambda (k_1 
\cdots 
k_n)}
+
 (\partial_{k_1}\cdots 
\partial_{k_n} \tilde{C}^{(3)}_{\mu \nu \lambda \rho \sigma }) 
\; I_4^{\mu \nu \lambda \rho \sigma (k_1 \cdots 
k_n)}\right] \nonumber
\end{eqnarray}
Here, all fields and their derivatives are evaluated at the origin of
9-dimensional space, around which point we have Taylor expanded the 
action.
$S_{{\rm flat}}$ is the flat space action for $N$ D0-branes, whose
leading terms are the dimensional reduction of D=10 SYM theory to 0+1
dimensions.  The complete form of the higher order terms in the flat
space action is not known, but the subleading terms vanish in the
matrix theory limit.  We assume that the background satisfies the
source-free IIA supergravity equations of motion so that the dual
fields $\tilde{C}, \tilde{B}, \tilde{C}^{(3)}$ are well-defined 7-, 6-
and 5-form fields given (at linear order) by
\begin{equation}
d \tilde{C} ={}^* dC, \;\;\;\;\;
d \tilde{B} ={}^* dB, \;\;\;\;\;
d \tilde{C}^{(3)} ={}^* dC^{(3)}.
\end{equation}
The sources $I_{2n}$ are associated with (multipole moments of)
Dirichlet $2n-$brane
currents, while the sources $I_{s}$ and $I_{5}$ are associated with
fundamental string and NS5-brane currents respectively.  

Starting from the general form (\ref{eq:IIA-general}), we applied the 
Seiberg-Sen 
prescription to arrive at an expression for the general background 
matrix theory action in terms of the unknown currents $I$. Demanding 
equivalence with our previously derived explicit results for the matrix 
theory action, we were led to a set of constraints on the currents $I$ 
which could then be solved to determine the leading (lowest dimension) 
term in each current (as well as some subleading terms) in terms of the 
matrix theory expressions $T$, $J$, and $M$ listed in the appendix.  
   The results for the lowest dimension operators appearing in the 
monopole (integrated) D0-brane currents are (with NS-NS currents in the 
left column and R-R currents in the right column)

\begin{equation}
\begin{array}{ll}
I_h^{00}  = T^{++} +  T^{+-} + (I_h^{00})_8 + {\cal O} (X^{12}) 
&I_0^0 = T^{++} \\
I_h^{0i}
=T^{+i} + T^{-i} + {\cal O} (X^{10}) &I_0^i = T^{+i} \\
I_h^{ij} = T^{ij} +
(I_h^{ij})_8 + {\cal O} (X^{12}) &
I_2^{ijk}  =  J^{ijk}+{\cal O} 
(X^8)\\
I_\phi = T^{++} -  \left({1 \over 3} T^{+-}  + {1
\over 3} T^{ii} \right) + (I_\phi)_8 + {\cal O} (X^{12}) &
I_2^{0ij}  =   J^{+ ij}  +{\cal O} (X^{10})
\\
I_s^{0i}  =  3J^{+ -i} +{\cal O} (X^8)&
I_4^{0i jkl}  =  6 M^{+ -i jkl}+{\cal O} (X^8) \\
I_s^{ij}  =  3 J^{+ ij}  -3J^{-ij}+
{\cal O} 
(X^{10})&I_4^{ij klm}  =  -6 M^{-ij klm}
+{\cal O} (X^{10})\\
I_5^{ijklmn}  =   {\cal O} (X^8)&
I_6^{0ijklmn}  =   S^{+ijklmn} +{\cal O} (X^{10}) \\
I_5^{0ij klm}  =   {\cal O} (X^{10}) &
I_6^{ijklmn p}  =   S^{ijklmn p}+{\cal O} (X^{12})
\end{array}
\label{eq:0-currents}
\end{equation}
while the higher moments $I_x^{(k_1 \cdots k_n)}$ of terms on the LHS
of each equation
are given by the higher moments of terms on the RHS.
The subleading terms mentioned explicitly in these operators, with 
dimensions denoted by a subscript, also satisfy the following relations, 
which may be derived from the agreement with matrix theory and the 
constraint of current conservation
\begin{eqnarray}
({1 \over 2} I_h^{00} + {1 \over 2} I_h^{ii}  + {3 \over 2} I_\phi)_8
& = &
T^{--} \nonumber\\
(I_h^{ij})_8 & = & (\partial_t T^{+i(j)} + \partial_t T^{-i(j)})_8
\nonumber\\
\end{eqnarray}
It is believed that the operators  $I_5^{\mu \nu \lambda \sigma \tau
\rho}$ representing NS5-brane current vanish identically; this issue
is discussed further in section 4.
There are additional terms analogous to those in
(\ref{eq:IIA-general}) describing the coupling to the fermionic
background fields of the IIA theory; these can be determined from the
results of \cite{Mark-Wati-3} by applying the Seiberg-Sen limiting 
procedure.

It may seem surprising that a system of D0-branes will couple to all
the supergravity fields $B_{\mu \nu}, C^{(3)}_{\mu \nu \lambda},
\ldots$ for which the usual sources are the fundamental string, the
D2-brane, etc.  In fact, however, this result follows from T-duality
from the well-known results that fundamental strings, momentum and
lower-dimensional D$(p -2k)$-branes can be described in terms of the
field strength of the $U(N)$ gauge field in the world-volume of a
D$p$-brane.  For example, the fact that in the system of D0-branes the
operator $J^{+ij}\sim[X^i, X^j]$ carries D2-brane charge is T-dual to
the statement that $F^{ij}$ flux on a D$p$-brane describes a
D$(p-2)$-brane extended in the directions perpendicular to the $i, j$
directions \cite{Douglas,WT-Trieste}.  In matrix theory this is
familiar as the mechanism by which a regularized membrane is composed
of $N$ pointlike partons \cite{Goldstone-Hoppe,BFSS}.  Similarly, the
statement that the operator $M^{+ -ijkl} \sim X^{[i} X^j X^k X^{l]}$
carries D4-brane charge \cite{grt,bss} is T-dual to the statement that
the instanton density $F\wedge F$ on a D$p$-brane corresponds to D$(p
-4)$-brane density \cite{Douglas}.  The results of \cite{Mark-Wati-4}
can be interpreted as giving explicit expressions for how the moments
of the currents associated with all higher-dimensional branes can be
expressed in terms of the matrix degrees of freedom of a system of
D0-branes.  The subset of these moments associated with conserved
brane charges were constructed using the supersymmetry algebra in
\cite{bss}.

We would like to emphasize that all the D0-brane operators determined 
explicitly by the comparison to matrix theory are defined in terms of a 
symmetrized trace over the non-abelian fields. However, it is not clear 
whether this prescription remains true for the higher dimension 
operators that we are unable to determine. 

\subsection{T-duality of supergravity backgrounds}

We would now like to T-dualize our results for the D0-brane case to find 
terms in the non-abelian Born-Infeld and Wess-Zumino actions for 
arbitrary D-branes which are linear in the background fields.  To do 
this, we must understand the 
action of T-duality both on the supergravity background fields and the 
fields living on the D-brane.  We discuss T-duality on the
background fields in this subsection and T-duality of the D-brane
field theory in the following subsection.

The usual T-duality of string theory on a compact circle can be
generalized to arbitrary backgrounds which are independent of the
compactification direction \cite{aal,bho}.  We are only interested at
this point in the linear parts of the T-duality transformation.  We
begin, however by recalling the complete transformation rules for a
set of background fields which are independent of a set of $n$
toroidally compactified coordinates $x^\ab$ on which we perform the
T-duality transform.  We will use the nonlinear form of the T-duality
transformation rules in Section 3 when we discuss the nonabelian
Born-Infeld action.  Throughout this section we use barred indices
$\ab, \bar{\beta}, \ldots$ for the compact spatial directions in which
a T-duality is performed and indices $\mu, \nu, \ldots$ for the
remaining $10-p$ space-time dimensions (including 0).

It turns out that the complete T-duality rules are greatly simplified  
if we combine the background fields into the combinations in which they
appear in the Born-Infeld and Wess-Zumino actions.  Thus, we define
fields 
\begin{eqnarray*}
{\cal G}_{\mu \nu} &=& g_{\mu \nu} - B_{\mu \nu}\\
{\cal C}^{(n)} &=& C^{(n)} - C^{(n-2)} \wedge B + {1 \over 2!} C^{(n-
4)} 
\wedge B \wedge B + \dots
\end{eqnarray*}
We now suppose that the fields are independent of $n$ coordinates
$x^{\bar{\alpha}}$ and apply the rules in \cite{bho} in order to 
T-dualize 
in  
each of the directions.  For the fields we have just defined, the 
complete non-linear transformation rules are simply\footnote{The 
transformation laws for the RR fields have additional terms in the case 
of massive type IIA supergravity.}
\begin{eqnarray}
{\cal G}_{\mu \nu} &\rightarrow& {\cal G}_{\mu\nu} - {\cal 
G}_{\mu\ab}{\cal G}^{\ab\bb}{\cal G}_{\bb\nu}\nonumber \\
{\cal G}_{\mu \ab} &\rightarrow& {\cal G}^{\ab\bb}{\cal 
G}_{\mu\bb}\nonumber \\
{\cal G}_{\ab\mu} &\rightarrow& -{\cal G}_{\bb\mu} {\cal 
G}^{\bb\ab}\nonumber \\
{\cal G}_{\ab\bb} &\rightarrow& {\cal G}^{\ab\bb}\label{eq:Tdual} \\
\phi &\rightarrow& \phi - {1 \over 2}{\rm ln}({\rm 
det}({\cal G}_{\ab\bb}))\nonumber \\
{\cal C}^{(q)}_{\mu_1 \dots \mu_{q-k} \ab_1 \dots \ab_k}
&\rightarrow& {1 \over (n-k)!} \epsilon^{\ab_1 \cdots \ab_{n}} 
{\cal C}^{(q-2k+p)}_{\mu_1 \dots \mu_{q-k} \ab_{k+1} \dots \ab_{n}}
 \nonumber
\end{eqnarray}
Here, ${\cal G}^{\ab\bb}$ is the inverse of ${\cal G}_{\ab\bb}$, so we 
see that under T-duality, the matrix ${\cal G}^{\ab\bb}$ which we have 
assumed to be constant transforms into its inverse.  

In this section
we will only use the linearized forms of these T-duality
relations.  Writing $g_{\mu \nu} = \eta_{\mu \nu} + h_{\mu \nu}$,
these read
\begin{eqnarray}
h_{\mu\nu} & \rightarrow & h_{\mu\nu}  \nonumber\\
b_{\mu\nu} & \rightarrow & b_{\mu\nu}  \nonumber\\
h_{\mu\ab} & \leftrightarrow &  -b_{\mu\ab}  \nonumber\\
h_{\ab\mu} & \leftrightarrow &  b_{\ab\mu}  
\label{eq:T-duality-linear}\\
h_{\ab\bb} & \rightarrow & -h_{\ab \bb}  \nonumber\\
b_{\ab\bb} & \rightarrow & -b_{\ab\bb}  \nonumber\\
\phi & \rightarrow & \phi -\frac{1}{2} \sum_{\ab} h_{\ab\ab} \nonumber\\
C^{(q)}_{\mu_1 \dots \mu_{q-k} \ab_1 \dots \ab_k}
&\rightarrow& {1 \over (n-k)!} \epsilon^{\ab_1 \cdots \ab_{n}} 
C^{(q-2k+p)}_{\mu_1 \dots \mu_{q-k} \ab_{k+1} \dots \ab_{n}}
 \nonumber
\end{eqnarray}

\subsection{T-duality of D-brane fields}

We now turn to the action of T-duality for the fields living on the
brane.  The low-energy action for a D$p$-brane living in flat space is
simply the dimensional reduction of $D=10$ SYM theory to $p+1$
dimensions.  If we retain the $D=10$ notation for all of the D$p$-brane
theories and simply reinterpret the expressions appropriately, (i.e.
$F_{ai}=D_aX^i$, $F_{ij}=i[X^i,X^j]$) then the expressions for the
low-energy actions are identical in each case.  In other words, the
action of T-duality on expressions written in terms of $D=10$
quantities is merely an appropriate reinterpretation of the notation.
The only subtlety in this interpretation arises when we consider the
transverse fields $X^i$ associated with a compact direction.  In this
situation, these fields are formally described by infinite matrices
containing $N \times N$ blocks indexed by an integer associated with
string winding number.  The components of these matrices are T-dual to
the momentum modes of the corresponding gauge field component $A_i$ on
the T-dual brane.  The details of this correspondence are worked out
in \cite{WT-compact}.

The matrix theory expressions for the moments of the $D=11$
supergravity currents may all be rewritten naturally in $D=10$
language.  Throughout the remainder of the paper we use these
expressions with the understanding that they should be interpreted in
terms of the appropriate $(p + 1)$-dimensional SYM variables in the
$p$-brane action.

\subsection{T-duality results: The D$p$-brane action }

Rewriting the known terms in the D0-brane action using the $D=10$
expressions for the matrix theory supercurrent components and
transforming using
the linearized T-duality rules  (\ref{eq:T-duality-linear}), it is a
fairly straightforward task to find the terms of interest
in all of the higher D$p$-brane actions.

Starting from the D0-brane action (\ref{eq:IIA-general}) and applying 
the rules above,
we find that the D$p$-brane action may be written\footnote{To write the
action in the Einstein frame, we make the substitution $h_{\mu \nu}
\rightarrow h_{\mu \nu} + {1 \over 2} \eta_{\mu \nu} 
\phi$ }${}^{,}$\footnote{Here, we use hatted indices for worldvolume 
spatial
indices (i.e.  excluding 0), while indices $i, j, \ldots$ denote
transverse spatial indices.}
\junk{ \begin{eqnarray}
S^{Dp} =& & ( \phi - {1 \over 2} h_{\hat{a} \hat{a}})(T^{++} - {1 
\over 3}T^{+-} - {1 \over 3}T^{ii} - {1 \over 3}T^{\hat{b}\hat{b}} + 
(I_{\phi})_8 + \cdots)\nonumber \\
&+& h_{00}(T^{++} + T^{+-} + (I_h^{00})_8 + \cdots)\nonumber \\
&+& {1 \over 2}h_{ij}(T^{ij} + (I_h^{ij})_8 +\cdots)\nonumber \\
&-& {1 \over 2}h_{\hat{a} \hat{b}}(T^{\hat{a} \hat{b}} + (I_h^{\hat{a} 
\hat{b}})_8 +\cdots)\nonumber \\
&+& h_{0i}(T^{+i} + T^{-i} + \dots)\nonumber \\
&+& h_{\hat{a}i}(6 J^{+ \hat{a} i } - 6 J^{- \hat{a} i} + 
\dots)\nonumber \\
&-& h_{0 \hat{a}}(6 J^{+-\hat{a}} + \dots) \label{eq:TJMaction} \\
&+& B_{ij}(3 J^{+ij} - 3 J^{-ij} + \dots)\nonumber \\
&-& B_{\hat{a} \hat{b}}(3 J^{+\hat{a} \hat{b}} - 3 J^{-\hat{a} \hat{b}} 
+ \dots )\nonumber \\
&+& B_{0i}(6J^{+-i} + \dots )\nonumber \\
&+& B_{\hat{a} i}(T^{\hat{a}i} + (I_h^{\hat{a} i})_8+ \dots)\nonumber \\
&+& B_{\hat{a} 0}(T^{+ \hat{a}} + T^{- \hat{a}} + \dots)\nonumber \\
&+& \{ {\rm higher \; moment \; terms} \} \nonumber
\end{eqnarray}
(*)ADD EXTRA $I_n$ TERMS(*)}
\begin{eqnarray}
S^{{\rm D}p}_{\rm NS-NS} & = & ( \phi - {1 \over 2} h_{\hat{a}
\hat{a}})I_\phi 
+\frac{1}{2} h_{00}I_h^{00}
+ {1 \over 2}h_{ij}I_h^{ij}
- {1 \over 2}h_{\hat{a} \hat{b}} I_h^{\hat{a} \hat{b}}
+ h_{0i}I_h^{0i}
+ 2h_{\hat{a}i} I_s^{\hat{a}i}
-  2h_{0 \hat{a}}I_s^{0 \hat{a}} \nonumber \\
&&+ B_{ij} I_s^{ij}
- B_{\hat{a} \hat{b}}I_s^{ \hat{a} \hat{b}}
+2 B_{0i} I_s^{0i}
+ B_{\hat{a} i}I_h^{\hat{a}i}
+ B_{\hat{a} 0}I_h^{0 \hat{a}}\label{eq:TJMaction}\\
& &
+ \{ {\rm higher \; moment \; terms} \} 
+ \{ {\rm nonlinear \; terms} \} 
\nonumber\\
\nonumber\\
S^{{\rm D}p}_{\rm R-R} & = &
\sum {(2n + p - q)! \over (n+p-q)! \; (q-n)! \; n!} \epsilon^{\ha_1 
\cdots 
\ha_p} C^{(q)}_{i_1 
\cdots i_n \ha_{n+p-q+1} \cdots \ha_p} I_{2n+p-q-1}^{i_1 \cdots i_n 
\ha_1 \cdots \ha_{n+p-q}}
\label{eq:general-R-R}\\& & + \sum {(2n + p - q +2)! \over 
(n+p-q+1)! \; (q-n-1)! 
\;n!} \;
\epsilon^{\ha_1 \cdots \ha_p} \;C^{(q)}_{i_1 \cdots i_n 0 \ha_{n+p-q+2} 
\cdots \ha_p} I_{2n+p-q-1}^{i_1 \cdots i_n 0 \ha_1 \cdots \ha_{n+p-q+1}}
\nonumber\\& &
+ \{ {\rm higher \; moment \; terms} \} 
+ \{ {\rm nonlinear \; terms} \} 
\nonumber
\end{eqnarray}
where $q, n$ are even/odd in the IIA/IIB theory.

The currents in these expressions are written in terms of the 0-brane
currents $I_x$ given in (\ref{eq:0-currents}) which in turn are given
in terms of the $T$'s, $J$'s and $M$'s listed in the appendix, and are
to be interpreted as $D=10$ expressions dimensionally reduced to $p+1$
dimensions.  The higher moment terms will be of exactly the
same form as the terms listed above, with arbitrary derivatives of each 
background field coupling to the appropriate higher moment of the 
corresponding current. For example, the lowest dimension terms coupling 
to $h_{ij}$ will be
\beas
& &{1 \over 2 n!} \partial_{k_1} \cdots \partial_{k_n} h_{ij} T^{ij (k_1 
\cdots k_n)}\\ &=& {1 \over 2 n!} \partial_{k_1} \cdots \partial_{k_n} 
h_{ij} \str((-D_a X^i D_a X^j - [X^i, X^k] [X^k,X^j]) X^{k_1} \cdots 
X^{k_n}) + \{{\rm fermions}\}
\eeas

As a consequence of our derivation from the D0-brane action, the index 0 
appears to be on a different footing than the remaining woldvolume 
indices $\hat{a}, \hat{b}, \dots$ in the expressions  
(\ref{eq:TJMaction}) and (\ref{eq:general-R-R}). For example, 
$B_{\hat{a} 
\hat{b}}$ couples to
\[
-I_s^{ \hat{a} \hat{b}} = -3 J^{+ \hat{a} \hat{b}} + 3 J^{- \hat{a} 
\hat{b}} + \dots
\]
while $B_{0 \hat{a}}$ couples to 
\[
-I_h^{0 \hat{a}} = -T^{+ \hat{a}} - T^{- \hat{a}} + \dots
\]
On the other hand, we would expect the D$p$-brane action to exhibit 
$p+1$ 
dimensional Lorentz symmetry in the worldvolume directions, therefore, 
$B_{ab}$ (including 0 indices) should couple to a Lorentz tensor 
$I^{ab}$. If so, it must be the case that
\beas
-3 J^{+ \hat{a} \hat{b}} + 3 J^{- \hat{a} \hat{b}} + \dots &=& 
I^{\hat{a} \hat{b}}\\
-T^{+ \hat{a}} - T^{- \hat{a}} + \dots &=& 2 I^{0 \hat{a}}
\eeas
for some tensor $I^{ab}$. Examining the expressions for $T$ and $J$ in 
the appendix, we see that the $D=10$ notation makes it clear that 
$T^{+\hat{a}} + T^{-\hat{a}}$ and $J^{+\hat{a}\hat{b}} - 
J^{-\hat{a}\hat{b}}$ are indeed $(0 \hat{a})$ and $(\hat{a}\hat{b})$ 
components of a Lorentz tensor. In a similar way, we find that $T^{ij}$ 
is related to $J^{+-i}$, and $J^{ijk}$ is related to 
$M^{+-ijkl}$. This was not obvious from the expressions for $T$, $J$, 
and $M$ as they appeared in previous papers. As a result of these 
relationships, 
it turns out that both expressions (\ref{eq:TJMaction}) and 
(\ref{eq:general-R-R}) are in fact Lorentz 
invariant. This provides a first check of our results, and fixes
several signs in the currents which were undetermined in 
\cite{Mark-Wati-3}.

Another obvious test of the results we have derived is to compare with 
the case $N=1$, where our results should match with the known abelian 
D$p$-brane actions. In particular, we note that for the case $p=9$, the 
world volume theory contains no scalar fields, and all of the terms we 
have written are symmetrized traces of expressions involving only the 
world-volume field strength, the fermions, and covariant derivatives on 
the fermions. As a result, if we specialize to the abelian U(1) case for 
$p=9$, the action will look the same as in the nonabelian theory, apart 
from covariant derivatives becoming usual derivatives and the ordering 
in products of field strengths $F_{\mu \nu}$ which must be taken into 
account in the nonabelian theory. 
This similarity between the terms in the $U(1)$ and $U(N)$ actions for 
the 9-brane is in contrast to all of the lower-dimensional brane actions 
which have terms containing $F_{ij} = i[X^i, X^j]$ and $D_i \Theta = i 
[X^i, \Theta]$ which vanish in the abelian theory. 
However, all of these terms T-dualize to terms in the $p=9$ action which 
do not vanish in the abelian case. Therefore a comparison of our $p=9$ 
results with the $p=9$ abelian Born-Infeld and Wess-Zumino actions will 
provide a check not only of the complete set of terms in our $p=9$ 
action, but also of strictly non-abelian terms in $p<9$ actions
\footnote{Since we must assume that the background fields are 
independent of directions transverse to the brane in order to T-dualize 
to $p=9$, we will not be able to check terms coupling to transverse 
derivatives of the background fields in this way}.
We perform this check 
in section 3.1, and in section 3.2, we will reverse this logic and use 
the full $p=9$ abelian Born-Infeld and Wess-Zumino actions to deduce a 
large 
set of additional terms in the non-abelian $p$-brane actions for $p<9$. 
In 
preparation for these comparisons, and as an example of the results from 
this section, we now write explicitly the terms we have derived for the 
case $p=9$.

\junk{

\subsection{T-duality results: R-R sector}

It is equally straightforward to apply the linearized T-duality rules
\cite{eq:T-duality-linear} to the R-R terms in the 0-brane action
(\ref{eq:IIA-general}).  For the terms in the general nonabelian
D$p$-brane
action coupling linearly to the background R-R fields we find
\begin{equation}
S^{{\rm D}p}_{\rm R-R} =
\sum {2n + p - q \choose n} \epsilon^{\ha_1 \cdots \ha_p} C^{(q)}_{i_1 
\cdots i_n \ha_{n+p-q+1} \cdots \ha_p} I_{2n+p-q-1}^{i_1 \cdots i_n 
\ha_1 \cdots \ha_{n+p-q}} + \sum {(2n + p - q +2)! \over (n+p-q+1)! n!} 
\epsilon^{\ha_1 \cdots \ha_p} C^{(q)}_{i_1 \cdots i_n 0 \ha_{n+p-q+2} 
\cdots \ha_p} I_{2n+p-q-1}^{i_1 \cdots i_n 0 \ha_1 \cdots \ha_{n+p-q+1}}
\label{eq:general-R-R}
\end{equation}
where $q, n$ are even/odd in the IIA/IIB theory.
As before, the currents $I_{2k}$ are the 0-brane R-R currents
given in (\ref{eq:0-currents}) which in turn are given
in terms of the $T$'s, $J$'s and $M$'s listed in the appendix, and are
to be interpreted as $D=10$ expressions dimensionally reduced to $p+1$
dimensions.  
}

\subsection{Example: D9-brane}
  In the case $p=9$, there are no transverse indices $i$ so the story 
simplifies considerably. The terms of lowest dimension in the
nonabelian theory coupling linearly to the NS-NS and R-R background
fields are given by 
\begin{eqnarray}
S^{\rm D9}_{NS-NS} &=& \str\left\{ ( \phi- {1 \over 2} h_{cc})
(1 + {1 \over 4}F_{ab}F^{ab} - {i \over 2}\Theta \slash{D} \Theta + 
(I_\phi)_8 +
\cdots)\right. \nonumber \\ 
& & \hspace{0.2in} - {1 \over 2}h_{ab}(F^{b c}F_c {}^a -
{i \over 2} \bar{\Theta} \Gamma^a D^b \Theta + (I_h^{ab})_8 +\cdots) 
\label{eq:D9_NS}\\
& & \hspace{0.2in} + {1 \over 2} B_{ab}(F^{ab} +  F^{a c} F_{c d} 
F^{d b} + {1 \over 4}F^{ab} F_{c d} F^{c d} \nonumber \\
& & \hspace{0.3in} - {i \over 4} F_{cd} \bar{\Theta} \Gamma^b 
\Gamma^{cd} D^a \Theta - {i \over 4} F_{cd} \bar{\Theta} \Gamma^d 
\Gamma^{ba} D^c \Theta + {i \over 8} D^d F_{dc} \bar{\Theta} 
\Gamma^{cab} \Theta +\cdots) \nonumber 
\end{eqnarray}
and\footnote{To derive the terms below by T-duality, we have assumed 
the background fields are constant, in which case the fermion terms are 
all total derivatives. These fermion terms do not seem to appear in the 
$\kappa$-symmetric actions, so it seems that we should integrate by 
parts before restoring spacetime dependence of the background.}
\begin{eqnarray}
S^{\rm D9}_{R-R} &=& \int d^{10} \sigma \epsilon^{abcdefghij}  \str[ 
c_{10}C^{(10)}_{abcdefghij}
 + c_{8} C^{(8)}_{abcdefgh}F_{ij} \nonumber\\
& &\hspace{1.5in} + c_{6}C^{(6)}_{abcdef}(F_{gh} F_{ij} + i\bar{\Theta} 
\Gamma^{ghi} D_j \Theta) \nonumber\\
& & \hspace{1.5in} + c_{4}C^{(4)}_{abcd}(F_{ef} F_{gh} F_{ij} + 
iF_{ef} \bar{\Theta} \Gamma^{ghi} D_j \Theta) \label{eq:D9_RR}\\
& & \hspace{1.5in} + c_{2}C^{(2)}_{ab}(F_{cd} F_{ef} F_{gh} F_{ij} 
+ F_{cd} 
F_{ef} i\bar{\Theta} \Gamma^{ghi} D_j \Theta +
{\cal O} (\Theta^4) ) \nonumber\\
& & \hspace{1.5in} + c_{0} C^{(0)}(F_{ab} F_{cd} F_{ef} F_{gh} 
F_{ij} + \{ 
{\cal O} (\Theta^2) ) ] \nonumber
\end{eqnarray}
where \footnote{The definition of $c_{2m}$ here reflects a change from 
(\ref{eq:general-R-R}) to more standard normalization conventions for 
the R-R 
fields.}
\[
c_{2m}  \equiv  {1 \over (2m)! (5-m)! 2^{5-m}} \nonumber
\]
Note that in order to T-dualize in all 9 spatial directions we have
assumed that the background fields are independent of all spatial
coordinates, so there are no higher moment contributions to this
action.  Of course, we expect that spatial variations in the
background fields can be incorporated into this action by simply
assuming that all world-volume fields as well as background fields in
the action have spatial dependence.  We will discuss this issue in
some further detail in section \ref{sec:Born-Infeld}.

\section{Comparison with the abelian theory}

In this section, we compare our results to the known abelian D-brane 
theories. In section (3.1), we show that the actions we have derived 
agree with a gauge fixed version of the abelian $\kappa$-symmetric 
D-brane actions. In section (3.2), we show that the abelian D9-brane 
action may be T-dualized to determine non-abelian terms in the lower 
D-brane actions. This provides an alternate derivation of some of our 
results plus a large set of additional higher order terms for both the 
Born-Infeld and Wess-Zumino actions.

\subsection{Abelian $\kappa$-symmetric D-brane theory}

We now compare our results to the known abelian $\kappa$-symmetric
action for a D$p$-brane. For a detailed discussion of these actions, see 
the original references \cite{aps1,aps2,cgnw,cgnsw,Bergshoeff-Townsend}.

   The $\kappa$-symmetric D$p$-brane actions are most naturally 
understood 
in terms of the superspace formalism for type IIA/IIB supergravity. In 
addition to the usual bosonic coordinates, type IIA/IIB superspace 
contains fermionic coordinates which form a pair of Majorana-Weyl 
spinors of opposite/the same chirality. 

The complete action may be written as the sum of a Born-Infeld type 
action 
and a Wess-Zumino action which take their usual form, 
\bea
S_{DBI} &=& - \int d^{p+1} \sigma e^{- \phi} \sqrt{-{\rm det} (g_{ij} + 
{\cal F}_{ij})} \nonumber\\
S_{WZ} &=& \int e^{\cal F} \sum_q C^{(q)}
\label{eq:DBI_WZ}
\eea
except that now the background fields are replaced by fields and 
differential forms on superspace. Here, we define
\[
{\cal F}_{ij} = F_{ij} - B^{wv}_{ij} 
\]
where $F_{ij}$ is the usual field strength of the world volume gauge 
field but now $B^{wv}_{ij}$ is the pullback of the superspace two-form 
$B_{MN}$ to the worldvolume. In the Wess-Zumino term, the sum runs 
over 
even RR-forms in the IIB case and odd RR-forms in the IIA case, which 
are also defined as pull-backs of superspace forms.
There are additional curvature terms which must appear in 
(\ref{eq:DBI_WZ})
to cancel anomalies in the world-volume theory
\cite{ghm,Minasian-Moore,Cheung-Yin}.  These terms are linear in the
background R-R fields and at least quadratic in curvature so they will
not affect our discussion of the linearized couplings and we will
ignore them here.

 The superspace geometry is described in terms of a super-vielbein one 
form 
\[
E^A = dZ^M E_M^A 
\]
and connection one form\footnote{We use $M,N, \dots$  
superspace coordinate indices, and $A,B, \dots$ for local frame 
indices. We will use lower case Latin indices $a,b,m,n, \dots$ refer to 
bosonic coordinates and lower case Greek indices to refer to fermionic 
coordinates. } $\omega^A_B$.  In terms of these, the pullback of the 
metric is given by
\[
g_{ij} =  \partial_i Z^M E_M^a \partial_j Z^N E_N^b \eta_{ab} 
\]
The fermionic components of the various forms are determined by 
constraints on their field strengths, which may be found in 
\cite{cgnsw}. These 
constraints are simplest in the basis defined by the one-forms $E^A$, 
however it is important to write the components with all bosonic indices 
in the coordinate basis rather than the frame basis so that the 
supergravity fields appearing in the final action are the usual ones. 
For example, we have in the type IIB case,
\beas
B^{wv}_{ij} &=& \partial_j Z^N E_N^B \partial_i Z^M E_M^A B_{AB}\\
&=& (\partial_j Z^n E_n^b + \partial_j Z^\nu E_\nu^b)  (\partial_i Z^m 
E_m^a + \partial_i Z^\mu E_\mu^a) B_{ab}\\
& & + 2 (\partial_j Z^n E_n^\beta + \partial_j Z^\nu E_\nu^\beta)  
(\partial_i Z^m E_n^a + \partial_i Z^\mu E_\mu^a) B_{a \beta}\\
& & + (\partial_j Z^n E_n^\beta + \partial_j Z^\nu E_\nu^\beta)  
(\partial_i Z^m E_n^\alpha + \partial_i Z^\mu E_\mu^\alpha) B_{\alpha 
\beta}\\
&=& (\partial_j X^n - i e_b^n \bar{\Theta} \Gamma^b \partial_j \Theta)  
(\partial_i X^m  - i e_a^m \bar{\Theta} \Gamma^a \partial_i \Theta) 
B_{mn}\\
& & + i (\bar{\Theta} \sigma_3 \Gamma_a \partial_j \Theta)  \partial_i 
X^m e_m^a   - i (\bar{\Theta} \sigma_3 \Gamma_a \partial_i \Theta)  
\partial_j X^m e_m^a \\ 
& & + {1 \over 2} (\bar{\Theta} \Gamma^a \partial_i \Theta) 
(\bar{\Theta} \Gamma_a  \sigma_3 \partial_j \Theta) - {1 \over 2} 
(\bar{\Theta} \Gamma^a \partial_j \Theta) (\bar{\Theta} \Gamma_a 
\sigma_3 \partial_i \Theta)\\
\eeas
Here we have used the fact that the superspace constraints on the 
torsion may be solved by
\[
\ba{cccc} E_m^a = e_m^a, & E_m^\alpha = 0, & E_\mu^a = 
i\delta_\mu^\alpha ( \Gamma^a \Theta)_\alpha, & E_ \mu^ \alpha 
= \delta_\mu^\alpha. \ea
\]
while the constraints on the field strength of $B$ may be solved by
\[
\ba{cc} B_{a \beta} = -i(\Gamma_a \sigma_3 \Theta)_\beta, & B_{\alpha 
\beta} =  -(\Gamma^a \Theta)_{(\alpha}(\Gamma_a \sigma_3 
\Theta)_{\beta)}
\ea
\]
The matrix $\sigma_3$ is a $2 \times 2$ Pauli matrix which acts on the 
index labeling the two fermionic superspace coordinates.

The Born-Infeld and Wess-Zumino actions above are separately invariant 
under 32 global supersymmetries which are guaranteed by their superspace 
covariant form. In addition, the combination is invariant under a 
16-dimensional local fermionic $\kappa$-symmetry. 

To relate these $\kappa$-symmetric actions, which have two spacetime 
Majorana-Weyl spinors and ten spacetime coordinates, to our action which 
is written in terms of a single worldsheet Majorana-Weyl spinor and 
$9-p$ scalars, we must use $\kappa$-symmetry to set one linear 
combination of the fermionic coordinates to zero and use 
reparametrization invariance to identify the first $p+1$ spacetime 
coordinates with the $p+1$ worldsheet coordinates (the static gauge). 
Since the local $\kappa$-symmetry is 16-dimensional, it is completely 
used up in eliminating half of the fermion fields. 

The fermion terms in the resulting action will depend on which 
combination of the fermions we choose to set to zero. To reproduce the 
results we have derived in the previous section, we must therefore make 
a specific choice for the fixing of $\kappa$-symmetry, however it is not 
obvious what this choice should be. Let us focus on the case $p=9$. In 
this case, it seems that any Lorentz invariant choice would amount to 
setting
\[
\Theta_1 = a\Theta, \; \; \; \; \; \Theta_2 = b\Theta.
\]  
One natural choice, taking $(a,b) = (0,1)$ or $(a,b) = (1,0)$ was
demonstrated for the flat space action in \cite{aps2}. With this
choice, all terms in the Wess-Zumino action independent of the
background fields vanish, leaving only the Born-Infeld action. It may
be checked that this choice gives an action whose fermion terms do not
agree with our results. In particular, it gives a dimension 4 fermion
operator coupling to $B$, while in our results, $B$ couples only to
operators of dimension 2 and 6. As it turns out, the abelian terms in
the actions we have derived seem consistent with another natural
choice\footnote{The factor $1/\sqrt{2}$ is chosen to match the
normalization conventions used above.}, $a=b=1/\sqrt{2}$, that is, we
set
\[
\Theta_1 = \Theta_2 = {1 \over \sqrt2} \Theta.
\]
This is the $p=9$ version of ``Killing Gauge'' \cite{Kallosh-killing}. 

To implement this gauge choice, we simply replace 
\[
\identity, \; \sigma_1 \rightarrow 1, \; \;\;\;\; \sigma_2, \sigma_3 
\rightarrow 0
\]
in all fermion bilinears. 

 With our choice of gauge the expressions for $g_{ij}$ and 
${\cal F}_{ij}$ simplify to
\bea
g_{ij} &=& (e_i^a - i \bar{\Theta} \Gamma^a \partial_i \Theta) \; (e_j^b 
- i \bar{\Theta} \Gamma^b \partial_j \Theta)\; \eta_{ab} \nonumber\\
{\cal F}_{ij} &=& F_{ij} - B^{wv}_{ij} \label{eq:gFfix}\\
&=&F_{ij} - (\delta_j^n - i e_b^n \bar{\Theta} \Gamma^b 
\partial_j \Theta)  
(\delta_i^m  - i e_a^m \bar{\Theta} \Gamma^a \partial_i \Theta) B_{mn}  
\nonumber
\eea
We are now in a position to explicitly write down the gauge fixed forms 
of the actions (\ref{eq:DBI_WZ}) and compare with the abelian version of 
our results. For the bosonic terms, we find complete agreement between 
the two methods for all of the operators we have derived. Comparing the 
fermion terms, we find something unexpected. The gauge fixed 
$\kappa$-symmetric action matches our results precisely if we assume 
that the $p$-form fields appearing in the matrix theory derived 
expressions (\ref{eq:D9_NS},\ref{eq:D9_RR}) are 
to be identified with world-volume fields rather than spacetime fields. 
We do not understand why this should be the case, however we note that 
since the D9-brane fills spacetime, the distinction between worldvolume 
indices and spacetime indices is somewhat subtle here.

As an example of this, we compute the operator coupling to $B_{mn}$ in 
the gauge fixed action. From (\ref{eq:DBI_WZ}), we see that $B$ appears 
in the action only in the combination ${\cal F}$. Thus, the 
complete dependence on $B$ may be easily obtained from the flat space 
action by making the substitution $F_{ij} \rightarrow {\cal F}_{ij}$. 
Up to dimension 8, the background independent part of the action, 
including contributions from both the Born-Infeld and Wess-Zumino terms 
is\footnote{In computing the Wess-Zumino terms, the relevant fermion 
terms in the superspace components of the R-R field strength are
\beas
C^{(2n)}_{a_1 \cdots a_{2n-1} \alpha} &=& e^{-\phi} \; (\Gamma_{a_1 
\cdots a_{2n-1}} (-\sigma_3)^n \sigma_2 \Theta)_\alpha\\ C^{(2n)}_{a_1 
\cdots a_{2n-2} \alpha \beta} &=& -ie^{-\phi} \; (\Gamma^{a_{2n-1}} 
\Theta)_\alpha (\Gamma_{a_1 \cdots a_{2n-1}} (-\sigma_3)^n \sigma_2 
\Theta)_\beta + \dots
\eeas
The omitted terms give no contribution with our choice of gauge, while 
the remaining components only contribute terms of dimension greater than 
8. Only the components of $C^{(4k+2)}$ give a non-vanishing contribution 
 in our gauge.} 
\beas
S_{D9} &=& -{1 \over 4} F_{ab}F_{ab} + {i \over 2} (\bar{\Theta} 
\slash{\partial} \Theta)\\
& & + {1 \over 8}(F_{ab} F_{bc} F_{cd} F_{da} - {1 \over 4} F_{ab} 
F_{ab} F_{cd} F_{cd}) + {i \over 16} (\bar{\Theta} \slash{\partial} 
\Theta) F_{ab} F_{ab} + {i \over 4} (\bar{\Theta} \Gamma^a \partial_b 
\Theta) F_{bc} F_{ca}\\
& & + {i \over 32} \bar{\Theta} \Gamma^{abcde} \partial_e \Theta F_{ab} 
F_{cd} + {1 \over 16} (\bar{\Theta} \slash{\partial} 
\Theta)(\bar{\Theta} 
\slash{\partial} \Theta) - {1 \over 16} (\bar{\Theta} \Gamma^a 
\partial_b 
\Theta)(\bar{\Theta} \Gamma^b \partial_a \Theta) + \dots
\eeas
To match conventions with the results in section 2, we have made a field 
redefinition $\Theta \rightarrow 2\Theta$.

Making the substitution $F_{ij} \rightarrow F_{ij}-B^{wv}_{ij}$ and 
keeping terms linear in $B^{wv}$, we find that up to dimension 6 
operators, the terms coupling to $B^{wv}$ are\footnote{Here, we ignore 
terms which vanish by the equations of motion}
\beas
{1 \over 2} B^{wv}_{ab} &(& F_{ab} + F_{ac} F_{cd} F_{db} + {1 
\over 4} 
F_{ab} F_{cd} F_{cd}\\ & & + {i \over 2} (\bar{\Theta} \Gamma^c 
\partial_b \Theta) F_{ac} + {i \over 2} (\bar{\Theta} \Gamma^b
\partial_c \Theta) F_{ac} + {i \over 4} (\bar{\Theta} \Gamma^{abc} 
\partial_d \Theta) F_{cd} + {i \over 4} (\bar{\Theta} \Gamma^{acd} 
\partial_b \Theta) F_{cd} + \dots)
\eeas
This expression agrees precisely with the abelian version of our results 
(\ref{eq:D9_NS}) if we identify the $B$ appearing in that equation with 
$B^{wv}$. We would have expected that the two expressions should have 
been compared after making the further substitution 
\[
B^{wv}_{ij} \rightarrow  (\delta_j^n - i e_b^n \bar{\Theta} \Gamma^b 
\partial_j \Theta)  (\delta_i^m  - i e_a^m \bar{\Theta} \Gamma^a 
\partial_i \Theta) B_{mn}
\]
however, this has the effect of eliminating the term ${i \over 2} 
(\bar{\Theta} \Gamma^b \partial_c \Theta) F_{ac}$, so there is a single 
term disagreement if we try to identify the $B$ in (\ref{eq:D9_NS}) with 
the $B$ having spacetime indices. The same effect appears when comparing 
the operators coupling to the Ramond-Ramond fields. 

It is possible that a different method of fixing $\kappa$-symmetry and 
reparametrization symmetry would give a gauge fixed action in which the 
spacetime $B$ appeared as in expression (\ref{eq:D9_NS}), however, it is 
difficult to see how a change in the gauge fixing could affect only the 
single term without changing the others that already agree. We leave it 
as a puzzle to understand why agreement is obtained by comparison with 
the pulled-back world-volume $p$-forms rather than spacetime $p$-form 
fields 
in the D9-brane action. 

One could repeat this analysis for the cases $p<9$ by gauge fixing the 
other abelian $\kappa$-symmetric D$p$-brane actions and comparing with 
our 
results, however, as discussed above, the comparison for $p=9$ already 
provides a check for all terms in the lower D$p$-brane actions which do 
not contain derivatives of the background fields in directions 
transverse to the brane. In fact, since many terms for $p<9$ vanish in 
the abelian case but T-dualize to $p=9$ terms which do not vanish, a 
comparison with the lower dimensional $\kappa$-symmetric actions would 
actually be a less stringent check than the one we have already 
performed. Rather, we will now exploit the connection between terms in 
the abelian 9-brane action and non-abelian terms in the lower brane 
actions to deduce a large set of additional terms for these actions.

\subsection{Non-abelian terms from the abelian $p=9$ action} 
\label{sec:Born-Infeld}

In this subsection we compare our results derived from matrix theory
and the D0-brane action by T-dualizing in $p$ directions to the action
found by T-dualizing the Born-Infeld action of a D9-brane in $9-p$
directions.  A related discussion is given in the recent paper
of Myers \cite{Myers-dielectric}.

\subsubsection{Bosonic Born-Infeld terms}

Let us first consider the bosonic terms in the Born-Infeld action for 
the case $p=9$. If, following the proposal of Tseytlin for the flat 
space case, we simply define the non-abelian Born-Infeld action for 
D9-branes to be that set of terms not containing any commutators of 
$F$'s \footnote{this is analogous to the abelian restriction to terms 
not containing derivatives of $F$'s since we may rewrite 
$[F_{ab},F_{cd}] = D_{[a}D_{b]}F_{cd}$.} then the remaining action 
will 
be (by definition) a symmetrized trace over products 
of $F$'s with indices contracted in various ways.  As argued above,
the restriction to the abelian case will clearly give exactly the same 
set of terms, the only difference being that the symmetrized trace may 
be dropped, since all the $F$'s commute already.  Thus, we conclude that 
the complete set of terms in the non-abelian action not containing 
commutators of $F$'s should be obtained from the abelian case by 
imposing a symmetrized trace, that is
\be
\label{eq:BI9}
S = \int d^{10}x \str\left(e^{-\phi}\sqrt{-{\rm det}(g_{ab} - B_{ab} + 
F_{ab})} \right)
\ee
This is a simple generalization of the proposal by Tseytlin for the flat 
space case. Starting from this action, we may now apply the T-duality 
rules above to find a large set of terms in the lower D$p$-brane 
actions. 
The resulting actions include terms with explicit commutators 
$[X^i,X^j]$ and are therefore are a much less obvious generalization of 
the corresponding abelian actions than (\ref{eq:BI9}). 

In our case, dualizing to determine the other D$p$-brane 
actions is more complicated than for flat space, due to the presence of 
background supergravity fields, however, the simple form of T-duality 
rules derived above will allow us to proceed without much difficulty.  

   We choose $n$ directions labeled by $x^i$ over which the T-duality is 
to be performed and assume for now that the background fields are 
independent of these directions.  Using our results for the D0-brane 
action, we will later be able to reinstate dependence on these 
directions (which will be transverse to the brane).  To perform the 
T-duality, we first rewrite the D9-brane action in terms of the field 
${\cal G} = g-B$ defined above, distinguishing between indices 
$i,j,\dots$ in 
the directions to be dualized and indices $a,b,\dots$ in the 
non-dualized directions that will become transverse and world-volume 
indices respectively. We have
\begin{eqnarray*}
S &=& \int d^{10}x \str \left(e^{-\phi}\sqrt{-{\rm det}({\cal G}_{ab} + 
F_{ab})} 
\right)\\
&=& \int d^{10}x \str \left(e^{-\phi}\sqrt{-{\rm det}\left( \ba{cc} 
{\cal G}_{ab} + 
F_{ab} & {\cal G}_{aj} + F_{aj} \\ {\cal G}_{ib} + F_{ib} & {\cal 
G}_{ij} + F_{ij} \ea  
\right)} \right)\\
&=& \int d^{10} x \str \left( e^{-\phi}\sqrt{-{\rm det}({\cal G}_{ij} + 
F_{ij}) 
{\rm 
det} ({\cal G}_{ab} + F_{ab} - ({\cal G}_{ai} + F_{ai})({\cal G}_{ij} + 
F_{ij})^{-1}({\cal G}_{jb} + 
F_{jb}) )} \right)\\
\end{eqnarray*}
where in the last line, we have used the identity
\begin{eqnarray*}
{\rm det} \left( \ba{cc} A & B \\ C & D \ea \right) &=& {\rm det} 
\left\{ \left( \ba{cc} \identity & 0 \\ 0 & D \ea \right)\left( \ba{cc} 
A & B \\ D^{-1}C & \identity \ea \right) \right\}\\
&=& {\rm det} (D) {\rm det}(A-BD^{-1}C)
\end{eqnarray*}
We now dimensionally reduce the world volume fields to  $p+1$ dimensions 
and replace the background fields by their duals.  The resulting 
expression is
\begin{eqnarray}
\label{eq:BI}
S^{Dp}_{BI} &=& T_p \int d^{p+1}\sigma  \str \left\{ 
e^{-\phi}\left(-det[\delta_{ij} + {\cal G}_{ik} F_{kj}])\right.  \right.  
\\
& & \hspace{0.5in} \left.  \left.  det[{\cal G}_{ab} + F_{ab} - 
{\cal G}_{ai}{\cal G}^{ij}{\cal G}_{jb}  + ({\cal G}_{ai} + F_{ak}{\cal 
G}_{ki})({\cal G} + {\cal G}F{\cal G})^{ij}({\cal G}_{jb} + 
{\cal G}_{jl}F_{bl}) ] \right)^{1/2} \right\} \nonumber
\end{eqnarray}
Here, ${\cal G}^{ij}$ is the inverse of ${\cal G}_{ij} = g_{ij} - 
B_{ij}$ and the term 
$({\cal G} + {\cal G}F{\cal G})^{ij}$ 
is the inverse of $({\cal G}_{ij} + {\cal G}_{ik}F_{kl}{\cal G}_{lj})$.  
We also recall that 
$F_{ai} = D_a X^i$, $F_{ij} = i[X^i, X^j]$, and $F_{ab}$ is the usual 
world volume field strength.  
 
   In the absence of background fields, it may be checked that this 
action reduces to Tseytlin's non-abelian flat space action.  In the 
abelian case with background fields, the action becomes
\begin{eqnarray*}
S^{Dp}_{U(1)}  &=& T_p \int d^{p+1}\sigma e^{-\phi} \sqrt{-det({\cal 
G}_{ab} + 
{\cal G}_{ai}F_{bi} + {\cal G}_{ib}F_{ai} + {\cal G}_{ij}F_{ai} F_{bj} 
+ 
F_{ab})}\\
&=& T_p \int d^{p+1}\sigma e^{-\phi} \sqrt{-det({\cal G}_{ab} + {\cal 
G}_{ai} 
\partial_b X^i + {\cal G}_{ib}\partial_a X^i + {\cal G}_{ij}\partial_a 
X^i\partial_b 
X^j + F_{ab})}
\end{eqnarray*}
which is equivalent to the usual covariant action
\[
S^{Dp} = T_p \int d^{p+1} \sigma e^{-\phi} \sqrt{ -det({\cal G}_{\mu 
\nu} 
\partial_a X^{\mu} \partial_b X^{\nu} + F_{ab})}
\]
in the static gauge where we identify $X^a = \sigma^a$.

   By expanding (\ref{eq:BI}) and keeping only terms linear in the 
background supergravity fields, one may verify that the leading terms 
are exactly the bosonic terms derived in section 2. 
   \junk{It may be checked  to see whether this action reproduces our 
terms we 
derived earlier from our D0-brane action.  Thus we expand (\ref{eq:BI}), 
keeping 
only terms linear in the background supergravity fields.  Alternatively, 
and perhaps more simply, we may do this expansion in the D9-brane action 
then use the simple linearized T-duality rules to get the desired terms.  
Either way, the result is
\begin{eqnarray}
S^{Dp} = &&\str \left\{  (1 - \phi + {1 \over 2} h_{aa})(1 + {1 \over 
4}F_{\mu \nu}F_{\mu \nu} + \cdots)\right.  \nonumber \\
& & \hspace{0.2in} +{1 \over 2}h_{ab}(F_{b \mu}F_{\mu a} + 
\cdots)\nonumber \\
& & \hspace{0.2in} + h_{ia}(F_{ai} + F_{a \mu} F_{\mu \nu} F_{\nu i} 
+ 
{1 \over 4}F_{ai} F_{\mu \nu} F_{\mu \nu}  +\cdots)\nonumber \\
& & \hspace{0.2in} +{1 \over 2} h_{ij}(F_{j \mu}F_{\mu i} + \cdots) 
\label{eq:linear} \\
& & \hspace{0.2in} +{1 \over 2}B_{ab}(F_{ba} + F_{b \mu} F_{\mu 
\nu} 
F_{\nu a} + {1 \over 4}F_{ba} F_{\mu \nu} F_{\mu \nu}  
+\cdots)\nonumber 
\\
& & \hspace{0.2in} +B_{ia}(F_{a \mu}F_{\mu i} + \cdots)\nonumber \\
& & \hspace{0.2in} - \left.  {1 \over 2}B_{ij}(F_{ji} + F_{j \mu} F_{\mu 
\nu} F_{\nu i} + {1 \over 4}F_{ji} F_{\mu \nu} F_{\mu \nu}  +\cdots) 
\right\} \nonumber
\end{eqnarray}
By explicitly inserting the the matrix theory expressions for the 
supergravity currents into (\ref{eq:TJMaction}), it may be checked that 
the bosonic terms  associated with constant background fields
exactly agree between the two approaches. } 

It is interesting to compare how the matrix theory and DBI approaches
describe coupling to background fields with spatial dependence.
First consider those directions perpendicular to the $p$-brane.  From 
the matrix theory approach, we may have dependence of the background 
supergravity fields on these transverse directions, which are described 
by the higher moment terms in (\ref{eq:TJMaction}).  
We do not know how to explicitly T-dualize background fields
depending on the compact coordinate, so we cannot determine these
terms in the $p$-brane DBI action by T-dualizing from the 9-brane.
For the bosonic terms, however, it is easy to see 
that we may reproduce the dependence found from the matrix theory
approach in the action simply by taking the background fields to be 
functions of the world-volume coordinates and the transverse scalar 
fields and then formally expanding in the transverse scalars inside the 
symmetrized trace. For example
\begin{eqnarray}
\phi &=& \phi(\sigma_o,\dots,\sigma_p, X^{p+1}, \dots, X^9) 
\label{eq:scalars}\\
&=& \sum_{n=0}^{\infty} {1 \over n!} \{ \partial_{k_1} \cdots 
\partial_{k_n} \phi\} (\sigma_o,\dots,\sigma_p, 0, \dots, 0) X^{k_1} 
\cdots X^{k_n} \nonumber
\end{eqnarray}
It seems natural to assume that this prescription also applies to the 
full non-linear action (\ref{eq:BI}), so we interpret all background 
fields in this expressions to be expansions of the form 
(\ref{eq:scalars}). Such a prescription has also been verified in 
certain cases by explicit string theory scattering calculations 
\cite{Garousi-Myers}.   

In contrast to the situation for the transverse coordinates,
the Born-Infeld approach gives an action in which the background
fields have a natural dependence on the world-volume coordinates
$\xi^a$, but such a dependence cannot be included when T-dualizing
from the 0-brane action.  It is quite straightforward to extend the
results of Section 2 by making all D-brane fields and background
fields explicitly dependent on the world-volume coordinates.

It would be interesting to invert this chain of argument and ask what
we can learn from this discussion about the action of T-duality in the
presence of a background field which depends on the compact coordinate.
In the language of the world-volume theory of the D$p$-brane wrapped
around a torus $T^p$, dependence of the background fields on the
compact degrees of freedom is naturally incorporated in the spatial
dependence of the background fields and the world-volume fields.
Formally inverting the T-duality transformation \cite{WT-compact} to
get a theory of (formally infinite) D0-brane matrix degrees of
freedom, we find a nontrivial coupling between block components of the
D0-brane matrices and the Fourier modes of the original background
fields.  Whether such a background can be interpreted as simply a
nontrivial classical supergravity background for the D0-brane system
is not clear; this may be a more general class of backgrounds with no
classical description, if indeed these backgrounds are consistent.  We
leave a further investigation of this question to further work. 

\subsubsection{Bosonic Wess-Zumino terms}
We now adopt a similar method to
derive terms in the non-abelian D$p$-brane Wess-Zumino actions.

We again start from the Abelian 9-brane action, which in terms of the 
field ${\cal C}\equiv Ce^{-B}$ defined above may be written
\begin{eqnarray*}
S^{D9}_{WZ} &=& \int \sum_{m=0}^5 {1 \over (5-m)!} {\cal C}^{(2m)} 
\{\wedge F \}^{5-m}\\
&=&\int d^{10} \sigma \epsilon^{a_0 \cdots a_9} \sum_{m=0}^5 
{1 \over (2m)! (5-m)! 2^{5-m} } {\cal C}^{(2m)}_{a_1 \cdots a_{2m}} 
F_{a_{2m+1} a_{2m+2}} \cdots F_{a_{9} 
a_{10}}
\end{eqnarray*}
Here, the $F$'s are already symmetrized, so there is no ordering 
ambiguity in passing to the non-abelian version.  Using the T-duality 
rules for the background fields (\ref{eq:Tdual}), we then find that the 
D$p$-brane WZ action is   
\be
S^{Dp}_{WZ} = \int d^{p+1}\sigma \epsilon^{a_0 \cdots a_p} \sum_{q} 
\sum_{n=max(0,q-p-1)}^{min(q,9-p)} c^p_{q,n} \str \left\{ {\cal 
C}^{(q)}_{ 
a_0 \cdots a_{q-n-1} i_1 \cdots i_n} F^{n+(p-q+1)/2}_{( a_{q-n} \cdots 
a_p i_1 \cdots i_n )} \right\}
\label{eq:WZ}
\ee
where
\[
c^p_{q,n} \equiv {(-1)^{n(n-1)/2} (p + 2n - q)!! \over n! \; (q-n)! \; 
(n+p-q+1)! }.
\]
In (\ref{eq:WZ}), the indices in brackets are to be assigned pairwise to 
the product 
of $F$'s and then symmetrized over all possible orderings. As for the 
Born-Infeld action, the background fields should be taken as functions 
of the world volume coordinates with scalar matrices inserted in place 
of the transverse coordinates, in order to reproduce the bosonic higher 
moment terms derived from matrix theory.  
  Again, we may compare terms linear in the background fields from this 
action to the terms we have derived from the D0-brane results, and we 
find that both methods give a consistent answer \footnote{The 
normalization 
conventions in the R-R fields in (\ref{eq:general-R-R}) are slightly 
different 
from the usual ones used here but the relation between the two 
conventions 
may be easily read off by comparing (\ref{eq:general-R-R}) and 
(\ref{eq:WZ}) }.

For $p<9$, this action contains many terms involving $F_{ij}$ which do
not appear in the abelian case.  Specifically, while the abelian
action has couplings involving only R-R q-forms with $q<p+1$ the
non-abelian action contains terms involving all allowed R-R fields
(i.e.  q odd/even for p even/odd).  Physically, just as a given
$p$-brane can carry $p-2n$ brane charge measured locally by $(F_{[a_1
a_2} \cdots F_{a_{2n-1} a_{2n}]})$, a collection of $p$-branes can
carry (moments of) $p+2n$ brane charge measured by $\tr(F_{[i_1 i_2}
\cdots F_{i_{2n-1} i_{2n}]})$ (with insertions of $X$'s).  As
mentioned in Section 2.1, this is perhaps most familiar from the
matrix theory point of view, where membrane  (D2-brane) states and
longitudinal 5-brane (D4-brane) states may be constructed out of
zerobranes.

\subsubsection{Fermion terms from $p=9$}

As for the bosonic terms, one could use the $p=9$ abelian action to
deduce the symmetrized trace terms with background fields independent
of the transverse directions for the fermionic terms in all the
non-abelian $p$-brane actions by imposing a symmetrized trace on the
$p=9$ action and dimensionally reducing. However, while for the
bosonic terms we were able to restore dependence of the background
fields on the transverse directions simply by making them functions of
the matrices $X^i$, such a prescription does not give all of the
fermion terms coupled to transverse derivatives of the background
fields, as may be seen in specific cases. Some of these terms would
appear in the abelian $p$-brane actions and should be derivable by
considering the $\kappa$-symmetric $p$-brane action, but there is no
clear way to find terms with commutators apart from our methods in the
section 2. Thus, it is not possible to reproduce all of our results
simply by considering the abelian actions.

\section{D3-branes and S-duality}

The SL(2,Z) S-duality symmetry of type IIB string theory maps a
D3-brane into another D3-brane \cite{Schwarz-multiplet}. We would like
to see how this S-duality invariance is manifested in the multiple
D$p$-brane action we have derived in the case $p = 3$. For the case of
a single D3-brane, the combination of Born-Infeld and Wess-Zumino
actions have been shown explicitly to lead to equations of motion
which possess an S-duality invariance involving simultaneous
transformations of the background fields and the world-volume fields
on the D3-brane
\cite{Gibbons-Rasheed,Tseytlin-dual,Green-Gutperle-D3,appsdual}. Since
we do not have the full non-abelian D3-brane actions it will not be
possible to demonstrate the complete S-duality of the theory with many
D3-branes in a type IIB supergravity background. However, using the
linear couplings we have derived and the known transformation
properties of the supergravity fields, we will show that the
requirement of S-duality gives information about the transformation
properties of various operators on the world-volume.  We find a new
prediction for the S-dual of certain operators constructed from the
transverse scalar fields in the SYM theory, and we show that this
duality transformation helps resolve an outstanding puzzle regarding
the transverse 5-brane of matrix theory.

\subsection{S-duality and the linearized action}

We focus on the $Z_2$ subgroup of the S-duality group generated by the
transformation $\tau$
which exchanges the 
NS-NS and R-R two form fields.  Under this transformation, background 
fields of IIB string theory transform at linear order as\footnote{Note
that these transformation rules are written in Einstein frame.
Although we have for the rest of this paper been working in string
frame, the terms in the action which we consider in this section  are 
the
same in both frames so it is consistent to use the Einstein frame
transformation rules in combination with the terms in the action
produced by specializing (\ref{eq:TJMaction},\ref{eq:general-R-R}) to
the case $p = 3$.}
\begin{eqnarray*} 
\phi & \rightarrow & -\phi \\
C^{(0)} & \rightarrow & -C^{(0)} \\
B_{\mu \nu} & \rightarrow & -C^{(2)}_{\mu \nu}\\
C_{\mu \nu} & \rightarrow & B^{(2)}_{\mu \nu}\\
g_{ij} & \rightarrow & g_{ij}\\
C^{(4)} & \rightarrow & C^{(4)}
\end{eqnarray*}

In order to check that our action satisfies S-duality we write
explicitly some of the terms in the D3-brane action, beginning with the
coupling of the world-volume bosonic fields to the NS-NS and R-R scalar 
fields.  The action in the absence of background fields is
\begin{equation}
S_0 ={\rm STr}\;\left(
 -\frac{1}{4}  F^2  -\frac{1}{8}  ( F^4- \frac{1}{4}  (F^2)^2) + \cdots
 \right)
\label{eq:no-background}
\end{equation}
The couplings to the background scalars are
\begin{equation}
\phi \cdot
  \left({1 \over 4}F_{ab} F^{ab} + \cdots \right)
\label{eq:dilaton-coupling}
\end{equation}
and
\begin{equation}
C^{(0)} \cdot \left( {1 \over 2} F \wedge F + \cdots \right)
 \label{eq:axion-coupling}
\end{equation}
The couplings to the NS-NS and R-R 2-form fields are
\begin{eqnarray}
 &  & {1 \over 2}B_{ab} \cdot \str \left( F^{ab}  +\frac{1}{4}F^{ab}F^2 
+
F^{a \mu} F_{\mu \nu} F^{\nu b} + \cdots \right)
\nonumber\\
 &  & -B_{ai} \cdot \str \left(F^{a \mu} F_{\mu i} + \cdots \right)
\label{eq:NS-NS-2-couplings}\\
& & -{1 \over 2} B_{ij} \cdot \str \left( F_{ij}
+\cdots \right) 
\nonumber
\end{eqnarray}
and
\begin{eqnarray}
 &  &- {1 \over 4} \epsilon^{abcd} C^{(2)}_{ab} \cdot \str \left( F_{cd} 
+ \cdots \right)
\nonumber\\
 &  & {1 \over 2}\epsilon^{abcd} C^{(2)}_{ai} \cdot \str \left( F_{bc} 
F_{di} + 
\cdots
 \right)\label{eq:R-R-2-couplings}\\
& & {1 \over 4}\epsilon^{abcd} C^{(2)}_{ij} \cdot  \str \left(  F_{ab} 
F_{ci} F_{dj} - {1 \over 4} F_{ab} F_{cd} F_{ij} 
+
\cdots \right) \nonumber
\end{eqnarray}

In the abelian case, the field strength of the gauge field on the 
world-volume of the D3-brane transforms at lowest order by
\begin{equation}
F_{ab} \rightarrow {1 \over 2} \epsilon_{abcd} F^{cd}.
\label{eq:f-s}
\end{equation}
There are higher order corrections to this transformation law from the
Born-Infeld action
\cite{Gibbons-Rasheed,Tseytlin-dual,Green-Gutperle-D3} which we will
not need to consider here.
{}From (\ref{eq:f-s}) it is straightforward to compute the
dual of the higher order terms in the field strength
\begin{eqnarray}
F_{ab}F^{ab} &  \rightarrow &  -F_{ab}F^{ab} 
 + \cdots \nonumber\\
F \wedge F & \rightarrow &  -F \wedge F + \cdots
\label{eq:more-abelian-S-duality}
\end{eqnarray}

We have no systematic knowledge of the S-duality transformation
properties of the world-volume operators in the nonabelian ${\cal N} =
4$ $U(N)$ gauge theory.
It is straightforward, however, to generalize the relations from the 
abelian
theory to analogous relations in the nonabelian theory by simply
performing a symmetrized trace on both sides of the relations
(\ref{eq:f-s}, \ref{eq:more-abelian-S-duality}).  
This gives
\begin{eqnarray}
{\rm STr}\;F_{ab} &  \rightarrow &
{\rm STr}\;\left( \frac{1}{2}\epsilon_{abcd} F^{cd} 
 +\cdots \right) \nonumber\\
{\rm STr}\;\left(F_{ab}F^{ab} \right)
 &  \rightarrow & {\rm STr}\; \left( - F_{ab}F^{ab} 
 + \cdots \right)\label{eq:nonabelian-S-duality}\\
{\rm STr}\;\left(F \wedge F \right)
 & \rightarrow &  {\rm STr}\;\left(-F \wedge F + \cdots \right)\nonumber
\end{eqnarray}
Note that in the nonabelian theory the duality transformation of the
higher order terms in $F$
are not automatic consequences of the duality
transformation of the  linear term.

Although the full abelian Born-Infeld action is not invariant under
S-duality, the equations of motion of this theory are invariant
\cite{Gibbons-Rasheed,Tseytlin-dual,Green-Gutperle-D3}.
We also expect that the terms describing the leading operators
coupling linearly to the background fields should transform among
themselves under S-duality.  The simplest argument for this is that we
would expect a system of D-branes which couple to a given background
supergravity field such as $B$ through an operator ${\cal O}_B$ to
couple to the S-dual supergravity field $\tilde{B}$ (in this case
$-C^{(2)}$) through the S-dual operator $\tilde{{\cal O}}_B={\cal
O}_{\tilde{B}}$.  In the AdS/CFT context a similar argument was given
in \cite{Intriligator}.

{}From the proposed nonabelian S-duality relations
(\ref{eq:nonabelian-S-duality}), it is straightforward to check that
the leading terms in the action
(\ref{eq:dilaton-coupling},\ref{eq:axion-coupling})
which are linear in the background
dilaton and axion fields transform into themselves under S-duality.
Note that we have so far avoided discussing the terms depending on the
world-volume scalars $X^i$.  The duality transformation properties of
the operators containing these fields are not known, although we will
now proceed to show that something can be learned about how these
operators transform from other terms in our action.

We now turn to an analysis of the S-duality properties of the terms
coupling to the background NS-NS and R-R 2-form fields.
The lowest dimension operators coupling to $B_{ab}$ and $C^{(2)}_{ab}$
are indeed related by nonabelian S-duality according to the
prescription above.  While the terms coupling to $B_{ai}$ and
$C^{(2)}_{ai}$ do not seem to be exactly S-dual, the leading operators
at least have the same dimension; we will return to these operators
momentarily.

When we consider the operators coupling to the 2-form fields $B_{ij}$
and $C^{(2)}_{ij}$ with purely transverse polarization indices, the
S-duality of our action seems at first glance to break down
completely.  There is a dimension 2 operator coupling to the NS-NS
2-form field, while the first operator coupling to the
R-R field has dimension 6.  In order to resolve this apparent
contradiction, we need to recall that each of these fields is related
to a dual 6-form field to which the NS-NS and R-R 5-branes are
electrically coupled.  In particular, consider the R-R 2-form field.
This field is related to the R-R 6-form field through
\begin{equation}
\partial_a C^{(2)}_{bcdefg} = {1 \over 3!} \epsilon_{abcdefgklm} 
\partial^k (C^{(2)})^{lm} 
\label{eq:26dual}
\end{equation}
Taking into account operators coupling to the R-R 6-form, which may be 
read off from (\ref{eq:general-R-R}), we find that the complete coupling 
of a
derivative of the transversely polarized R-R 2-form field is 
\begin{equation}
 \partial_k C^{(2)}_{ij}  \left( -{1 \over 12}\epsilon_{ijklmn} {\rm 
Tr}\; (X^l [X^m, X^n])
+ {1 \over 4}{\rm Tr}\; (X^k \epsilon^{abcd} (F_{ab} F_{ci} F_{dj} -{1 
\over 4} F_{ab} F_{cd} F_{ij})) + \cdots \right)
\label{eq:leading-t-R-R}
\end{equation}
Let us compare this to the terms coupling to the transversely
polarized NS-NS 2-form field.  Since ${\rm Tr}\;[A, B]$ vanishes for
any $U(N)$ matrices $A, B$, the first nonvanishing term appears in the
first moment and is given by
\begin{equation}
-{1 \over 2}\partial_k B_{ij} \left( {\rm Tr}\; (X^k[X^i, X^j]) + \cdots 
\right)
\label{eq:dbterm}
\end{equation}
If we assume that S-duality acts on the
combination of transverse fields ${\rm Tr}\; ([X^i, X^j] X^k)$ through
\begin{equation}
S:{\rm Tr}\; ([X^i, X^j] X^k) \rightarrow
-{1 \over 6} \epsilon^{ijklmn} {\rm Tr}\; ([X^l, X^m] X^n).
\label{eq:XXXtrans}
\end{equation}
then the term (\ref{eq:dbterm}) is precisely the S-dual of the leading
operator in (\ref{eq:leading-t-R-R}).  Thus, we see that the condition
that our action is S-duality invariant can be satisfied and indeed
may give new information about how operators formed from the transverse
fields transform under S-duality.

The mechanism we have just discovered can be seen to solve 
other apparent problems with the S-duality of the couplings
(\ref{eq:NS-NS-2-couplings}) and (\ref{eq:R-R-2-couplings}).  As noted
above, the dimension 4 operators coupling to the fields $B_{ai}$ and
$C^{(2)}_{ai}$ are not completely equivalent under S-duality.  In
particular, the operator coupling to the NS-NS field contains terms of
the form $F_{ai} F_{ij}$ while the operator coupling to the R-R field
contains no such terms.  However, taking into account contributions from 
the 6-form using (\ref{eq:26dual}) we see that the complete coupling of  
derivative of this background field is\footnote{We do not need to worry 
about the constant pieces since constant $B_{ai}$ and $C_{ai}$ can be 
gauged away.}
\begin{equation}
\partial_jC^{(2)}_{ai}\cdot \str \left( {1 \over 2} \epsilon^{abcd} 
F_{bc} F_{di}  X^j + {1 \over 6}
\epsilon^{ijklmn}F_{al} F_{mn} X^k + \cdots \right)
\end{equation}
which has the correct S-duality relation with the term coupling to
$\partial_jB_{ai}$ as long as (\ref{eq:XXXtrans}) continues to hold with 
an insertion of $F_{ai}$. 

We have thus seen that the S-duality invariance of our linearly
coupled action can be explained by S-duality relations on the
operators appearing in the maximally symmetric super Yang-Mills theory
such as (\ref{eq:XXXtrans}).  It is interesting to compare this result
with the conjecture of Intriligator \cite{Intriligator} regarding the
S-duality transformation properties of short operators in the SYM
theory in the context of the AdS/CFT correspondence.  The operators in
the SYM theory are characterized by their transformation properties
under the superconformal symmetry group $PSU (2, 2 | 4)$.  The
operators which linearly couple to the background supergravity fields
live in short representations of the superconformal algebra, both in the
AdS/CFT correspondence and also in the action we are discussing here.
(We will discuss the AdS/CFT correspondence and its connection to our
action in more detail in the next section.)
The operators in short representations of $PSU(2, 2 |4)$
were classified in
\cite{Gunaydin-Marcus}, and an explicit table listing which operators
in the SYM theory correspond to which representations is given in
\cite{Intriligator}.  In general, the short operators are uniquely
determined by their $SO(4)$ and  $SU(4)$ transformation properties, and
must therefore transform into themselves under S-duality up to a sign.
In \cite{Intriligator}, a conjecture was made for the modular weights
of each of the short operators in the theory.
The operators considered in (\ref{eq:XXXtrans}) can be combined into
self-dual and anti-self-dual combinations under the 6-dimensional
duality transformation arising from contraction with
$\epsilon^{ijklmn}$.  Each of these combinations lives in an irreducible
representation of $SO(6) \approx SU(4)$.  Our prediction that
\begin{equation}
S:{\rm Tr}\; ([X^i, X^j] X^k) \rightarrow
-{1 \over 6} \epsilon^{ijklmn} {\rm Tr}\; ([X^l, X^m] X^n)
\label{eq:XXXtrans-2}
\end{equation}
corresponds to the statement that the self-dual and
anti-self-dual combinations of these operators under the 6-dimensional
epsilon symbol also transform into themselves under S-duality so that
one combination is self-S-dual and the other is anti-self-S-dual.

\subsection{S-Duality and transverse 5-branes in matrix theory}

We conclude this section with a brief description of how this
discussion of S-duality relates to the problem of the transverse
5-brane in matrix theory\footnote{Thanks to Ofer Aharony for
discussions related to this issue}.  
Consider a
configuration of the D3-brane theory on a 3-torus with transverse
fields $X^4, X^5$ and $X^6$ given by
\begin{equation}
X^{3 + i} = rJ^i
\end{equation}
where $J^i$ are the generators of the N-dimensional representation of
SU(2).  This configuration contains a D5-brane with the geometry $T^3
\times S^2$ where $S^2$ is a sphere of radius $r$ in the space spanned
by $X^4-X^6$.  Under T-duality on the $T^3$ this becomes the standard
membrane sphere of matrix theory \cite{Dan-Wati}.  If the D5-brane
configuration is acted on by S-duality, the resulting configuration
contains a NS5-brane of geometry $T^3 \times S^2$.  The T-dual of this
is an NS5-brane constructed from D0-branes, which should carry the
first moment of the NS5-brane charge $I_5^{012345(6)}$, and which
should couple (magnetically)
to the first derivative of the NS-NS 2-form field of the
IIA theory.  But we do not know any expression for the NS5-brane
charge of a system of 0-branes.  Such an expression would correspond
to a charge for the transverse 5-brane of matrix theory,
which is believed to vanish identically.

This apparent puzzle can be resolved by observing that even if the
moment $I_5^{012345(6)}$ of the NS5-brane charge vanishes, the
NS5-brane configuration we have just described couples correctly to
the 2-form field of the IIA theory.  The initial configuration carried
a nonzero charge ${\rm Tr}\; (X^{[4} X^5 X^{6]})$.  The S-dual IIB
NS5-brane configuration thus carries a charge ${\rm Tr}\; (X^{[7} X^8
X^{9]})$.  The same charge is carried by the T-dual IIA configuration
which should describe $N$ 0-branes as well as an NS5-brane with
geometry $T^3 \times S^2$.  In the IIA background-dependent D0-brane
action (\ref{eq:IIA-general}) this charge couples to the field
$\partial_{[7}B_{89]}$.  But this field is dual to the component
$\partial_{[6} \tilde{B}_{012345]}$ which we expect to describe the
appropriate dipole moment of the NS5-brane configuration.  Thus, we
see that although there is no operator describing transverse
NS5-branes in matrix theory, and no corresponding NS5-brane operator
$I_5^{0ijklm}$ in the IIA theory, it is still possible to construct
0-brane configurations which describe finite volume NS5-branes with
higher multipole moments, which couple correctly to the background
gravitational fields. In matrix theory, such compact transverse M5-brane 
configurations will couple correctly to the supergravity 3-form field
even though the operator $M^{+ijklm(n)}$ is identically zero.
   
\section{Relation to the AdS/CFT correspondence}

In this section we discuss applications of our results to the study of 
D-brane black holes and a relation to the $AdS/CFT$ correspondence.

\subsection{Absorption by black holes}

The most obvious application of the actions we have derived is to the 
study of D-branes interacting with weak bulk supergravity fields. For 
example, the  probability  for absorption of a particle $\phi$ by a 
stack 
of coincident branes is exactly determined by the operators coupling 
linearly to $\phi$ in the nonabelian D$p$-brane action. If $\phi$ 
couples 
to an operator $O_\phi$, then the absorption cross-section for $\phi$ 
may be read off from the two point function (see 
eg.\cite{Gubser-Klebanov})
\[
\langle O_\phi(x) O_\phi(0) \rangle
\]
  At low energies, only the lowest dimension part of the operator will 
contribute, and these lowest dimension parts have been completely 
determined by our results for all supergravity fields. A specific 
example which has already appeared \cite{ktv}, is the study of dilaton 
absorption by D3-branes. In that paper, we used our results to determine 
the lowest dimension operators coupling to all derivatives of the 
dilaton field in the nonabelian D3-brane action. Using these operators, 
we were able to compute the low-energy absorption cross-section for all 
partial waves of the dilaton. We found that the results precisely agreed 
with the classical supergravity calculation for dilaton partial wave 
absorption in the D3-brane geometry, giving evidence for the equivalence 
between the D3-brane world-volume theory and physics of supergravity 
near 
the branes. 

The exact numerical agreement found provided a stringent check of our 
results here, including normalizations. In particular, we note that the 
symmetrized trace ordering prescription was crucial for agreement. An 
important point to note about the comparison is that the supergravity 
calculation is reliable only in a limit corresponding to strong t'Hooft 
coupling in the world-volume theory, which ensures weak curvatures in 
supergravity. The agreement with our gauge theory calculation, carried 
out at weak coupling, implies that the calculated two point functions 
are not renormalized. Using the operators we have derived, it should be 
possible to perform an explicit gauge theory calculation showing that 
the first $g^2N$ correction vanishes. This would be an excellent test of 
our results, since the ``purely nonabelian'' terms in our operators, 
involving commutators of nonabelian fields, contribute to the two point 
function only at this subleading order.  

\subsection{Particle-operator correspondence}

Particle absorption calculations of the type just discussed provided
early motivation for the AdS/CFT conjecture
\cite{Maldacena-conjecture}, which in the most well-known example
states that the physics of supergravity in the near-horizon geometry
of D3-branes ($AdS^5 \times S^5$) is exactly described by the low
energy physics of the D3-brane world-volume fields, namely $N=4$, $D=4$
Super-Yang-Mills theory. In this subsection, we describe how the
actions we have derived may be used to obtain useful information about
the AdS/CFT correspondence, and also how the AdS/CFT correspondence
may be used to obtain useful information about matrix theory.

Consider again the process of particle absorption by D3-branes.  At
very low energies, the near horizon region and the asymptotically flat
part of the D3-brane geometry almost decouple. In this limit, particle
absorption may be understood as a particle $P_0$ in the asymptotic
region producing a very weak excitation in the transition region
between Minkowski space and $AdS^5 \times S^5$ which in turn excites
one of the particles $P$ of $AdS^5 \times S^5$ (see for example
\cite{Das-Mathur-2}). In the D-brane world-volume picture, the particle
$P_0$ turns on an operator $O_P$ in the gauge theory which governs the
decay of $P_0$ into particles on the world-volume. The exact
correspondence between particles on $AdS^5
\times 
S^5$ and operators in the $N=4$ Super-Yang-Mills world-volume theory is 
precisely the correspondence between the particles $P$ and the operators 
$O_P$ which are excited by a given asymptotic particle $P_0$. 

This logic suggests a precise way of using our actions to determine the 
operator $O_P$ corresponding to a given supergravity particle $P$ in 
$AdS^5 \times S^5$, including normalization. We consider the full 
supergravity solution corresponding to a stack of a large number of D3-
branes in the limit where the near horizon geometry is almost decoupled.
As for $AdS^5 \times S^5$, this space will have a set of independently 
propagating excitations (normal modes). Very near the brane, the 
equations of motion reduce to those for $AdS^5 \times S^5$, so one of 
the particles (normal modes) $P'$ of the full supergravity solution must 
correspond to our particle $P$. In the asymptotically flat region, $P'$ 
will look like some freely propagating particle $P_0$. This is precisely 
the particle, discussed above, that when sent in from infinity will 
excite the particle $P$ in AdS. Thus, to determine the gauge theory 
operator corresponding to $P$, we simply read off the leading operator 
coupling 
to $P_0$ in the nonabelian D3-brane action expanded about flat space. 

Though this prescription is well defined in principle, it is usually a 
nontrivial task to determine the particle $P_0$ in the asymptotic 
region which corresponds to the AdS particle $P$. It requires 
diagonalizing the supergravity equations of motion in the two regions 
(and sometimes in an intermediate region) and matching the solutions on 
the overlaps. In particular, the combination of supergravity fields 
describing $P_0$ may in general be different than that describing the 
particle $P$ in the $AdS$ region. 

In \cite{Das-Trivedi}, Das and Trivedi gave a slightly different
prescription, also using the D3-brane action, for determining the
relevant operators.  Rather than choosing the operator coupling to
$P_0$ in the expansion of the flat space D3-brane action, they
suggested taking the operator coupling to the particle $P$ in the
D3-brane action expanded about AdS space.  It may be that both
prescriptions give the same result. Certainly in the case of a
minimally coupled scalar, the supergravity fields defining $P$ and
$P_0$ are the same, so the two methods will give the same operator at
least up to normalization.  For fields with more complicated
propagators, further work is needed to make the correspondence between
operators on the D-brane and asymptotic fields in the extremal
geometry precise.

\subsection{AdS/CFT and matrix theory}

Our results also suggest a precise link between matrix theory and the
AdS/CFT conjecture (for previous discussions of connections between
these conjectures, see
\cite{ 
Hyun,Chepelev-are,deAlwis-correspondence,Polchinski-M-Theory,Silva}).
In short, the $D=10$ Super-Yang-Mills theory operators whose
dimensional reduction to $D=3+1$ correspond to particles in $AdS^5
\times S^5$ are linear combinations of the operators whose dimensional
reductions to $D=0+1$ describe the 11-dimensional supergravity
currents corresponding to a given matrix theory configuration.  On the
AdS side, we have a series of operators corresponding to an infinite
tower of Kaluza-Klein modes for each particle, wheras in matrix
theory, the series of operators corresponds to the infinite set of
multipole moments of a given current\footnote{These matrix theory
operators have also been related to the AdS/CFT correspondence by
relating them to a tower of Kaluza-Klein states in the near-horizon
geometry of  $N$ D0-branes
\cite{Sekino-Yoneya}.}. An explicit example of this was given in
\cite{ktv}, where the Yang-Mills operator corresponding to the $l$-th
Kaluza-Klein mode of the dilaton was found to be a combination of the
$l$th multipole moment of various components of the Matrix Theory
stress-energy tensor\footnote{Here, the indices $i,k$ are transverse
to the brane, while the index $\hat{a}$ runs over the spatial brane
directions.}, \be O_{\phi}^{k_1 \cdots k_l} = {1 \over 6} T^{ii(k_1
\cdots k_l)} - {1 \over 3} T^{\hat{a} \hat{a} (k_1 \cdots k_l)} - {1
\over 3} T^{+- (k_1 \cdots k_l)}.
\label{eq:dilop}
\ee

One particularly interesting aspect of this equivalence is that whereas 
the matrix theory expressions for the supergravity currents were 
determined by a general a one-loop gauge theory calculation, the 
operators corresponding to particles in $AdS$ may be determined (except 
for normalization) simply by acting with tree level supersymmetry 
generators on the chiral primary operators
\be
\str(X^{i_1} \cdots X^{i_l}) - \{ {\rm traces}\}
\label{eq:CPO}
\ee
By the correspondence, it should therefore be possible to obtain the 
results of one-loop matrix theory calculations simply by computing the 
action of tree level supersymmetry generators on chiral primary fields. 
This observation was put into practice in \cite{ktv} where 
the four 
fermion terms in the operators corresponding to all partial waves of the 
dilaton field (and therefore in the matrix theory currents on the right 
side of (\ref{eq:dilop})) were computed
by acting with four supercharges on the chiral 
primary 
operators (\ref{eq:CPO}). The matrix theory operator 
corresponding to the integrated component $T^{--}$ of the stress energy 
tensor is essentially the complete one-loop matrix theory effective 
action, which has the form 
\[
\str(F^4 - \frac{1}{4}  (F^2)^2 + {\rm fermions}).
\]
On the AdS side, we recognize this as the highest dimension operator
$O_8$ corresponding to the Weyl mode of the metric. So a complete
calculation of the one-loop Super-Yang-Mills theory effective action
appears to be equivalent to simply applying eight supercharges to the
chiral primary operator $\str(X^iX^jX^kX^l)$.  We expect that the
agreement between these two approaches to calculating the dimension 8
operator is a result of the high degree of supersymmetry in the
system--indeed, another way to determine the structure of this
operator would be to generalize the methods of \cite{pss} to show that
in the $SU(N)$ theory this operator is uniquely determined by
supersymmetry.

Finally, we note that since all $N=4$ SYM operators corresponding to 
particles in $AdS^5 \times S^5$ are constrained by supersymmetry to be 
symmetrized (the short representations to which these operators belong 
are obtained by acting with supercharges on the symmetrized chiral 
primary operator), an equivalence between these operators and the matrix 
theory current operators would provide an explanation for the 
symmetrized trace both in the matrix theory operators and in leading 
terms in the D-brane actions which we have derived. In this case, the 
terms in the D$p$-brane actions which vanish in the matrix theory or 
AdS/CFT limits\footnote{Note that it appears to be the same set of 
terms which are preserved in both limits} may not be subject to the same 
constraints from supersymmetry and therefore might be expected to 
contain non-symmetrized terms. We discuss this point further below. 

\section{Discussion}

In this paper, we have derived the leading operators in all D$p$-brane 
actions coupling linearly to each supergravity field, 
(\ref{eq:TJMaction},\ref{eq:general-R-R}). Our results have been 
obtained by applying the rules of T-duality to previous results for the 
D0-brane action. Using the same T-duality rules we have shown that the 
abelian D9-brane action may be used to deduce a large set of additional 
terms in the symmetrized part of the nonabelian actions for all lower 
D$p$-branes. These expressions, which include terms with arbitrary 
numbers 
of background and world-volume fields appear as equations (\ref{eq:BI}) 
and (\ref{eq:WZ}) respectively. These expressions do not represent the 
complete D$p$-brane actions, and we now summarize the types of 
corrections 
that we expect to occur.     

\subsection{Higher order corrections}

 All of the leading operators that we have derived from matrix theory
results are written in terms of a symmetrized trace over the nonabelian
fields.  We have argued above that this symmetrized trace prescription
may be viewed as a constraint of supersymmetry on the terms in the
D-brane actions which survive in the matrix theory or AdS/CFT limits.
For higher order terms that vanish in these limits, the constraints no
longer apply, so we do not have any reason to believe that the
symmetrized trace prescription continues to hold.  In particular, the
higher order expressions (\ref{eq:BI}) and (\ref{eq:WZ}) derived from
the D9-brane action have been written in terms of a symmetrized trace
simply because we have ignored possible terms with commutators of $F$'s.
Indeed, it has been argued that such terms are necessary at order
$F^6$ in the flat space nonabelian Born-Infeld action \cite{Bain}.  
These commutator terms may be viewed as higher derivative
terms which vanish in the abelian theory, since $[F_{ab}, F_{cd}] =
D_{[a}D_{b]}F_{cd}$.  The D$p$-brane actions also contain explicit 
higher
derivative terms which do not vanish for the Abelian case.  These are
excluded in the definition of the Born-Infeld action; for a discussion 
see \cite{Tseytlin-review} and references therein.

A further type of correction to the actions we have derived involves
terms of higher order in the background fields. The expressions
(\ref{eq:BI}) and (\ref{eq:WZ}) derived from the D9-brane actions
contain terms at all orders in the background fields, however it may
be shown that these expressions are not complete since they do not
satisfy the ``geodesic length criterion'' proposed by Douglas 
\cite{Douglas-curved-2}.  This
criterion is relevant when we consider two parallel separated branes,
described in the world-volume theory by giving an expectation value to
the diagonal scalar fields which correspond to the positions of the
two branes. For non-zero separation, the off-diagonal scalar fields
which arise from strings stretching between the branes should acquire
a mass equal to the geodesic distance between the branes. At linear
order in the background metric, it was shown in
\cite{Mark-Wati-4} that this geodesic distance criterion is exactly 
satisfied for the 
D0-brane action we have found. However, for agreement at second order in 
the metric, it is possible to show explicitly that additional terms 
involving $(\partial h)^2$ are required beyond those appearing in the 
action (\ref{eq:BI}). Some of these terms have been worked out in the 
case of D-branes on Kahler manifolds in \cite{Douglas-curved, dko}, but
for the most part these terms are unknown.  Another set of terms which
appear at higher order in the background fields arise from the
anomalous couplings discussed in
\cite{ghm,Minasian-Moore,Cheung-Yin}.  These terms presumably all have
nonabelian counterparts which remain to be investigated.

Finally, we note that there may be terms in the 
D$p$-brane action involving more than a single trace over world-volume
fields. These would seem necessary in order to reproduce string theory
amplitudes whose wordsheet involved more than a single boundary, for
example an annulus diagram with various world-volume fields inserted
on each boundary. It is possible that some of these terms could be
derived using our methods by performing a matrix theory calculation at
higher than one loop.

We have listed all possible additional terms which may appear in a
nonabelian Born-Infeld type action in a general background, based on
our understanding of the fields and symmetries of the theory as well
as the idea that such an action is defined as a sum over string
backgrounds.  It is not completely clear that there is really a unique
well-defined supersymmetric nonabelian Born-Infeld action which will
satisfy all the criteria which the abelian Born-Infeld action
satisfies and which additionally satisfies Douglas' geodesic length
condition and the condition discussed in \cite{Hashimoto-Taylor} that
the string spectrum describing small fluctuations around a fixed
D-brane background should be reproduced by the NBI action.
The terms we have found here are good evidence that it is 
possible to extend the nonabelian Super-Yang-Mills theory describing
multiple D$p$-branes to include higher order terms.
One might argue, however, that these linear couplings between
background fields and the multiple D$p$-brane world-volume theory are
protected by supersymmetry and are therefore uniquely determined while
the definition of the higher order terms in the nonabelian Born-Infeld
action might be more ambiguous due to the lack of supersymmetric
protection.  It will be very interesting to see whether further work
will reveal a well-defined NBI action at all orders.

\subsection{Further directions}

The most obvious direction in which this work could be pursued further
is to try to use higher-order matrix theory calculations to extend the
Born-Infeld action to include terms coupling to higher powers of the
background fields.  The next concrete step would be to generalize the
3-graviton calculation of Okawa and Yoneya \cite{Okawa-Yoneya} to a
general 3-body calculation, which would be a general 2-loop
calculation in an $SU(N)$ gauge theory with a block-diagonal
background containing 3 blocks of arbitrary size.  As argued in
\cite{Mark-Wati-3}, this general result would indicate the structure
of the quadratic coupling of a matrix theory object to background
fields, and by using the Seiberg-Sen limit this result should
translate into a well-defined result for the quadratic couplings of a
system of  $N$ D0-branes to the background fields.  By using T-duality
as in this paper, this would lead to the quadratic couplings of a
system of D$p$-branes of arbitrary dimension to the background
fields.  The correspondence between the matrix theory calculation just
described and supergravity interactions at the first nonlinear order
depends upon the nonrenormalization of the general 2-loop $SU(N)$
calculation.  The agreement of the 2-loop calculation in the case of
$SU(3)$ with supergravity gives hope that such a nonrenormalization
theorem may hold for general $N$.  Note, however, that the methods of
\cite{pss} do not apply to the 2-loop $SU(N)$ calculation when $N > 3$
so the terms we need may not be renormalized \cite{Sethi-Stern-2}.  If
this is the case this approach may not help in understanding the
nonlinear terms in the NBI action until a better understanding is
found of how matrix theory behaves in the large $N$ limit.

There are many other problems to which the work described here can be
applied.  It would be nice to see how far the results of \cite{ktv}
can be generalized to describe absorption of particles by D$p$-branes
at leading and higher orders.  The relationship between T-duality and
spatial dependence of background supergravity fields is also an
interesting question for further investigation.  It would also be
interesting to carry out a more systematic analysis of the S-duality
of the linear terms we have found in the NBI action; this should give
a complete set of results for the S-duality transformation properties
of all the short operators in maximally supersymmetric 4D SYM theory.

\section*{Acknowledgments}

We would like to thank Ofer Aharony, Lorenzo Cornalba, Ken
Intriligator, Karl Millar,
Samir Mathur, Rob Myers, Ricardo Schiappa and John
Schwarz for helpful conversations.  We would like to thank Rob Myers
for sharing an early draft of his preprint \cite{Myers-dielectric}
with us.  The work of MVR is supported in part by the Natural Sciences
and Engineering Research Council of Canada (NSERC).  The work of WT is
supported in part by the A.\ P.\ Sloan Foundation and in part by the
DOE through contract
\#DE-FC02-94ER40818.

\newpage

\appendix

\section{Supercurrents from matrix theory}

We reproduce here for convenience the matrix theory forms of the
multipole moments of the 11D supercurrent found in
\cite{Dan-Wati-2,Mark-Wati-3}.  Dropping a factor of $1/R$ from each 
expression,
the stress tensor $T^{IJ}$, membrane
current $J^{IJK}$ and 5-brane current $M^{IJKLMN}$
have integrated (monopole) components

\begin{eqnarray*}
T^{++} &=& \str(\identity) = N\\
T^{+i} &=& -\str(F^{0i})\\
T^{+-} &=& \str(F^{0\mu}F^0{}_\mu + {1 \over 4} F_{\mu \nu} 
F^{\mu \nu} + {i \over 2} \bar{\Theta} \Gamma^0 D_0 \Theta)\\
T^{ij} &=& \str(F^{i \mu} F_\mu {}^j + {i \over 4} \bar{\Theta} 
\Gamma^i D_j \Theta + {i \over 4} \bar{\Theta} \Gamma^j D_i \Theta)\\
T^{-i} &=& -\str(F^{0 \mu} F_{\mu \nu} F^{\nu i} + {1 \over 4} 
F^{0 i} F_{\mu \nu} F^{\mu \nu}\\ 
& & \hspace{0.5in} - {i \over 8} F_{\mu \nu} \bar{\Theta} \Gamma^i 
\Gamma^{\mu \nu} D_0 \Theta + {i \over 8} F_{\mu \nu} \bar{\Theta} 
\Gamma^0 
\Gamma^{\mu \nu} D_i \Theta - {i \over 4} F_{\mu \nu}\bar{\Theta} 
\Gamma^\nu \Gamma^{0i} D^\mu \Theta\\
& & \hspace{0.5in} - {1 \over 8} \bar{\Theta} \Gamma^{0 \mu i} \Theta 
\bar{\Theta} \Gamma_{\mu}  \Theta)\\
T^{--} &=& {1 \over 4} \str(F_{\mu \nu} F^{\nu \gamma} 
F_{\lambda 
\delta} F^{\delta \mu} - {1 \over 4} F_{\mu \nu}F^{\mu 
\nu}F_{\gamma \delta}F^{\gamma \delta}\\
& & \hspace{0.5in} + i F_{\mu \nu} F_{\gamma \delta} \bar{\Theta} 
\Gamma^{\nu} \Gamma^{\gamma \delta} D^{\mu} \Theta + \{ {\rm 
four 
\; fermion \; terms} \} )\\
J^{+ij} &=& -{1 \over 6} \str(F^{ij})\\
J^{+-i} &=& {1 \over 6} \str(F^{0 \mu} F_\mu {}^i + {i \over 4} 
\bar{\Theta} \Gamma^0 D_i \Theta - {i \over 4} \bar{\Theta} \Gamma^i 
D_0 
\Theta)\\
J^{ijk} &=& +{1 \over 6} \str( F^{0i} F^{jk} + F^{0 j} F^{k i} + F^{0 
k} F^{ij} - {3i  \over 4} \bar{\Theta} \Gamma^{0 [i j} D^{k]} \Theta + 
{i \over 4} \bar{\Theta} \Gamma^{ijk} D_0 \Theta)\\
J^{-ij} &=& {1 \over 6} \str(F^{i \mu} F_{\mu \nu} F^{\nu j} 
+ 
{1 \over 4} F^{ij} F_{\mu \nu} F^{\mu \nu}\\ 
& & \hspace{0.5in} - {i \over 8} F_{\mu \nu} \bar{\Theta} \Gamma^j 
\Gamma^{\mu \nu} D_i \Theta + {i \over 8} F_{\mu \nu} \bar{\Theta} 
\Gamma^i \Gamma^{\mu \nu} D_j \Theta - {i \over 4} F_{\mu 
\nu}\bar{\Theta} \Gamma^\nu 
\Gamma^{ij} D^\mu \Theta\\
& & \hspace{0.5in} + {1 \over 8} \bar{\Theta} \Gamma^{ \mu i j} 
\Theta \bar{\Theta} \Gamma_{\mu} \Theta)\\
M^{+-ijkl} &=& {1 \over 12} \str(F^{ij} F^{kl} + F^{ik} F^{lj} + F^{il} 
F^{jk} - i\bar{\Theta} \Gamma^{[ijk}D^{l]} \Theta)\\
M^{-ijklm} &=& -{5 \over 4} \str(F^{0[i}F^{jk} F^{lm]} + {i \over 2} 
F^{[0i} \bar{\Theta} \Gamma^{jkl} D^{m]} \Theta)
\end{eqnarray*}

Here, $\str$ denotes a symmetrized trace in which we average over all 
possible orderings of the matrices the trace, with commutators being 
treated as a unit. Time derivatives are taken with respect to Minkowski 
time $t$. Indices $i, j, \ldots$ run from 1 through 9, while indices $a, 
b,\ldots$ run from 0 through 9.
In these expressions we have used the definitions $F_{0i} = \dot{X}^i,
F_{ij} = i[X^i, X^j]$.
We do not know of a matrix form for the transverse 5-brane current
components $M^{+ ijklm},M^{ijklmn}$, and in fact comparison with 
supergravity suggests that these should be 0 for any matrix 
theory configuration.

The higher multipole moments of these currents contain
one set of terms which are found by including the matrices $X^{k_1},
\ldots, X^{k_n}$ into the symmetrized trace as well as more
complicated spin contributions. We may write these as
\begin{eqnarray}
T^{IJ (i_1 \cdots i_k)} & = & \sym (T^{IJ}; X^{i_1}, \ldots, X^{i_k}) + 
T_{\rm fermion}^{IJ(i_1 \cdots i_k)} \nonumber\\
J^{IJK (i_1 \cdots i_k)} & = & \sym (J^{IJK}; X^{i_1}, \ldots, X^{i_k}) 
+ 
J_{\rm fermion}^{IJK(i_1 \cdots i_k)}\\
M^{IJKLMN (i_1 \cdots i_k)} & = & \sym (M^{IJKLMN}; X^{i_1}, 
\ldots, 
X^{i_k}) + M_{\rm fermion}^{IJKLMN(i_1 \cdots i_k)}\nonumber
\end{eqnarray}
where some simple examples of the two-fermion contribution to the
first moment terms are
\begin{eqnarray*}
T_{\rm fermion}^{+i(j)} &=& -{1 \over 8R} \tr(\bar{\Theta} 
\Gamma^{[0ij]}  \Theta)\\
T_{\rm fermion}^{+-(i)} &=& -{1 \over 16R} \tr(F_{\mu \nu}\bar{\Theta}  
\gamma^{[\mu \nu i]} \Theta -4\bar{\Theta} F_{0\mu} 
\gamma^{[0\mu i]} \Theta)\\
T_{\rm fermion}^{ij(l)} &=& {1 \over 8R} \tr(F_{j \mu}\bar{\Theta}  
\gamma^{[\mu il]} \Theta + \bar{\Theta} F_{i\mu} \gamma^{[\mu jl]} 
\Theta)\\
J_{\rm fermion}^{+ij(k)} &=& {i \over 48R} 
\tr(\bar{\Theta}\Gamma^{[ijk]} \Theta)\\
J_{\rm fermion}^{+-i(j)} &=& {1 \over 48R} \tr(F_{0\mu}\bar{\Theta}  
\gamma^{[\mu ij]} \Theta + \bar{\Theta} F_{i\mu} \gamma^{[\mu 0j]} 
\Theta)\\
M_{\rm fermion}^{+-ijkl(m)} &=& -{i \over 16R} \str\left(\bar{\Theta} 
F^{[jk}\Gamma^{il]m}\Theta\right)
\end{eqnarray*}  
The remaining two-fermion contributions to the first moments and some
four-fermion terms are also determined by the results in
\cite{Mark-Wati-3}.

There are also fermionic components of the supercurrent which couple
to background fermion fields in the supergravity theory.  We have not
discussed these couplings in this paper, but the matrix theory form of
the currents is determined in \cite{Mark-Wati-3}

There is also a 6-brane current appearing in matrix theory related to
nontrivial 11D background metrics.  The components of this current as
well as its first moments are
\begin{eqnarray}
S^{+ijklmn} & = & \frac{1}{R}  \str\left(F_{[ij} F_{kl} F_{mn]}\right) 
\nonumber\\
S^{+ijklmn(p)} & = & \frac{1}{R}  \str\left(F_{[ij} F_{kl} F_{mn]} 
X_{p} 
-\theta F_{[kl}F_{mn}\gamma_{pqr]}  \theta\right) 
\label{eq:6-current2}\\
S^{ijklmnp} & = & 
\frac{7}{R}  \str\left(F_{[ij} F_{kl} F_{mn} \dot{X}_{p]}+ 
(\theta^2, \theta^4 \; {\rm terms})\right) 
\nonumber\\
S^{ijklmnp(q)} & = & 
\frac{7}{R}  \str\left(F_{[ij} F_{kl} F_{mn} \dot{X}_{p]} X_{q}  
- \theta \, \dot{X}_{[j} F_{kl} F_{mn} \gamma_{pqr]} \theta +{i \over 2} 
\theta 
\,
 F_{[jk} F_{lm}  
F_{np} \gamma_{qr]} \theta
\right) \nonumber
\end{eqnarray}

\bibliographystyle{plain}

\end{document}